\documentclass[12pt]{article}
\usepackage{titling}
\usepackage{amsmath}
\usepackage{slashed}
\usepackage{amssymb}
\usepackage{epsfig}
\usepackage{graphicx}
\usepackage{multirow,booktabs}
\usepackage{tabu}
\usepackage{color}
\usepackage[normal]{subfigure}
\usepackage{rotating}
\usepackage{hyperref,afterpage}
\usepackage[margin=1.0in]{geometry}
\usepackage[usenames,dvipsnames,svgnames,table]{xcolor}
\usepackage{enumitem}
\usepackage[utf8x]{inputenc}
\usepackage[compress,numbers,sort]{natbib}
\usepackage{authblk}
\usepackage{colortbl}
\usepackage{pdflscape}
\usepackage{braket}
\usepackage{color}
 \usepackage{pgfplotstable}
\usepackage{array}
\usepackage{colortbl}
\usepackage{float}
\usepackage{siunitx}
\usepackage{placeins}
\definecolor{Gray}{gray}{0.95}
\newcommand{\CG}{\cellcolor{Gray}}

\usepackage{xparse}
\ExplSyntaxOn
\NewDocumentCommand{\longdash}{ O{2} }
{
	--\prg_replicate:nn { #1 - 1 } { \negthinspace -- }
}
\ExplSyntaxOff

\definecolor{nicered}{rgb}{0.6,0.1,0.1}
\definecolor{nicegreen}{rgb}{0.1,0.5,0.1}
\definecolor{mediumcandyapplered}{rgb}{0.99, 0.12, 0.07}
\definecolor{red}{rgb}{1.0, 0, 0}
\hypersetup{colorlinks,citecolor= nicegreen,linkcolor= nicered}

\def\eq#1{{eq.~(\ref{#1})}}
\def\eqs#1#2{{eqs.~(\ref{#1})--(\ref{#2})}}

\def\Re{\mbox{Re}\,}

\newcommand{\mr}[1]{\multirow{2}{*}{#1} }

\renewcommand{\bar}{\overline}

\newcommand{\hc}{\mbox{h.c.}}
\newcommand{\SM}{\textrm{SM}}
\newcommand{\LO}{\textrm{LO}}
\newcommand{\NLO}{\textrm{NLO}}

\definecolor{LightCyan}{rgb}{0.88,1,1}
\definecolor{piggypink}{rgb}{0.99, 0.87, 0.9}
\definecolor{applegreen}{rgb}{0.55, 0.71, 0.0}
\definecolor{darkpastelgreen}{rgb}{0.01, 0.75, 0.24}
\definecolor{green-yellow}{rgb}{0.68, 1.0, 0.18}

\newcommand{\beq}{\begin{equation}}
\newcommand{\eeq}{\end{equation}}
\newcommand{\bea}{\begin{eqnarray}}
\newcommand{\eea}{\end{eqnarray}}
\newcommand{\MSbar}{$\bar{\text{MS}}$}

\newcommand{\published}[1]{%
\gdef\puB{#1}}
\newcommand{\puB}{}

\title{
\bf{Higgs probes of top quark contact interactions and their interplay with the Higgs self-coupling}
}
\author[1]
{Lina Alasfar\thanks{alasfarl@physik.hu-berlin.de}}
\affil[1]{\emph{\normalsize Institut f\"ur Physik, Humboldt-Universit\"at zu Berlin, D-12489 Berlin, Germany}}
\author[2]
{Jorge de Blas\thanks{deblasm@ugr.es}}
\affil[2]{\emph{\normalsize CAFPE and Departamento de F\'isica Te\'orica y del Cosmos,
Universidad de Granada, Campus de Fuentenueva, E--18071 Granada, Spain}}
\author[3]
{Ramona Gr\"{o}ber\thanks{ramona.groeber@pd.infn.it}}

\affil[3]{\emph{\normalsize Dipartimento di Fisica e Astronomia ``G. Galilei", Universit\`a di Padova, and
Istituto Nazionale di Fisica Nucleare, Sezione di Padova,
I-35131 Padova, Italy}}

\date{}

\published{\flushright 	HU-EP-21/48-RTG \vskip1cm}

\begin{document} 

\maketitle

\begin{abstract}
\normalsize
We calculate the dominant contributions of third generation four-quark operators to single-Higgs production and decay.
They enter via loop corrections to Higgs decays into gluons, photons and $b\bar{b}$, and in Higgs production via gluon fusion and in association with top quark pairs. 
We show that these loop effects can, in some cases, lead to better constraints than those from fits to top quark data. 
Finally, we investigate whether these four-fermion operators can spoil the determination of the trilinear Higgs self-coupling from fits to single-Higgs data.
\end{abstract}

\clearpage

\section{Introduction}

The precise determination of the Higgs boson properties is one of the main focus of the Large Hadron Collider (LHC) physics programme. Within the current experimental precision, the measurements of the Higgs couplings so far appear to be in agreement with the Standard Model (SM) predictions within an accuracy of, typically, ten percent \cite{Aad:2019mbh, Sirunyan:2018koj}. In many beyond the SM (BSM) scenarios, however, it is expected that new physics will introduce modifications in the Higgs properties. 
If the new BSM degrees of freedom are much heavier than the electroweak scale, a general description of potential new physics effects can be formulated in the language of an effective field theory (EFT). One possibility of such a parameterization is the so-called Standard Model EFT (SMEFT), in which new physics effects are given in terms of higher-dimensional operators involving only SM fields and that also respect the SM gauge symmetries.  The dominant effects on Higgs physics, electroweak physics and top quark physics stem from dimension-six operators, suppressed by the new physics scale $\Lambda$. This approach is justified in the limit 
in which energy scales $E\ll \Lambda$ are probed. 
\par
In this paper we will consider a small subset of these operators, namely four-fermion operators of the third generation quarks.
A direct measurement of the four-top quark operators requires the production of four top quarks. At the LHC, for $\sqrt{s}=13\;\text{TeV}$, and within the SM, this is a rather rare process, with a cross section of about 12 fb including NLO QCD and NLO electroweak (EW) corrections \cite{Frederix:2017wme}. This is due to the large phase space required for the production of four on-shell top quarks. First experimental measurements \cite{Aad:2020klt} indicate a slightly higher cross section than the SM prediction.\footnote{We note that a CMS combination from different LHC runs \cite{Sirunyan:2019nxl}, though having lower signal significance, shows agreement with the SM prediction.} Though four-top production gives direct access to four-top operators, the main effect comes from $\mathcal{O}(1/\Lambda^4)$ terms when computing the matrix element squared \cite{Hartland:2019bjb}, questioning whether one should neglect, in general, the effects of dimension-eight operators in the calculation of the amplitudes. At any rate, current experimental bounds on the four-top operators are rather weak. A significant improvement in constraining power would be expected, however, at a future 100 TeV $pp$ collider, due to the growth with the energy of the diagrams involving four-top operators \cite{Banelli:2020iau}. 
The situation is rather similar for the operators leading to $t\bar{t}b\bar{b}$ contact interactions. They can be measured directly in $t\bar{t}b\bar{b}$ production, see \cite{Sirunyan:2020kgar, ATLAS:2018gug} for experimental analyses at $\sqrt{s}=13\text{ TeV}$, but also leading to rather weak limits in SMEFT fits \cite{DHondt:2018cww, Hartland:2019bjb}.
\par
Given the rather weak ``direct'' bounds on the $t\bar{t}t\bar{t}$ and $t\bar{t}b\bar{b}$ contact interactions, here we will discuss alternative probes, showing how these interactions can be constrained indirectly via their contributions to single-Higgs observables.\footnote{Alternatively, other indirect probes of four-top quark interactions that have been proposed include top quark pair production \cite{Degrande:2020evl} and electroweak precision data \cite{deBlas:2015aea,Dawson:2022bxd}. The latter mostly leads to bounds on operators that can be constrained only weakly from Higgs data.}
These operators generate contributions to the effective couplings of the Higgs to gluons and photons via two-loop diagrams. At the one-loop level, they also modify associated production of a Higgs boson with top quarks and, in the case of $t\bar{t}b\bar{b}$ operators, also the Higgs decay to bottom quarks. While the leading log results can be easily included by renormalisation group operator mixing effects \cite{Jenkins:2013zja, Jenkins:2013wua, Alonso:2013hga}, in this paper we will compute also the finite terms and show that they can be numerically important.
\par
In addition, we will study the interplay between the extraction of the Higgs self-coupling measurement from single-Higgs production and decay and the four-fermion operators.
It was previously proposed that competitive limits 
to the ones from Higgs pair production on the trilinear Higgs self-coupling can be set using single-Higgs data \cite{McCullough:2013rea, Gorbahn:2016uoy, Degrassi:2016wml, Bizon:2016wgr, Maltoni:2017ims, Degrassi:2019yix, Degrassi:2021uik, Haisch:2021hvy}.  A global fit including all operators entering in Higgs production and decay at tree-level plus the loop-modifications via the trilinear Higgs self-coupling has been performed in \cite{DiVita:2017eyz}. 
Searches for modifications of the trilinear Higgs self-coupling via single-Higgs production have been presented by the ATLAS  \cite{ATLAS:2019pbo} and CMS \cite{CMS:2020gsy} collaboration. Using the example of the four-quark operators, we will show that there are other weakly constrained dimension-six  operators, that enter at the loop level, that should be included in such analyses as they have a non-trivial interplay with the trilinear Higgs self-coupling extraction from single-Higgs measurements. We will hence perform a series of combined fits of these four-fermion operators together with the operator modifying the trilinear Higgs self-coupling. While our study does not consider a global fit to all operators entering Higgs data, the results of our computations can be easily used in global analyses. Our main point, namely that in a global fit all operators entering via loop contributions, if so far constrained only weakly (as it is the case for, e.g., four-top operators), should be included, is clearly demonstrated by our few-parameter fits. 
\par 
The paper is structured as follows: in~\autoref{sec:notation} we introduce the notation used for the effective Lagrangian in our analysis. In~\autoref{sec:Higgs} we give the results of our computation of the loop contributions of the four-fermion operators. The results of our fits to Higgs data including the computed loop contributions are presented in~\autoref{sec:fit}, 
where we show results for both current data and projections at the High-Luminosity LHC (HL-LHC).
We conclude in~\autoref{sec:conclusion}. 
Further details of our analysis and additional material derived from our results are presented in two appendices.


\section{Notation \label{sec:notation}}

In the presence of a gap between the electroweak scale and the scale of new physics, $\Lambda$, the effect of new particles below the new physics scale can be described by an EFT. In the case of the SMEFT, the SM Lagrangian is extended by a tower of higher-dimensional operators, ${\cal O}_i$, built using the SM symmetries and fields (with the Higgs field belonging to an $SU(2)_L$ doublet), and whose interaction strength is controlled by Wilson coefficients, $C_i$, suppressed by the corresponding inverse power of $\Lambda$. In a theory where baryon and lepton number are preserved, the leading order (LO) new physics effects are described by the dimension-six  SMEFT Lagrangian,
\begin{equation}
\mathcal{L}_{\mathrm{SMEFT}}^{d=6}=\mathcal{L}_{\SM} + \frac{1}{\Lambda^2}\sum_i C_i  {\cal O}_i.
\end{equation}
A complete basis of independent dimension-six operators was presented for the first time in \cite{Grzadkowski:2010es}, the so-called {\it Warsaw basis}. In this work, we are interested in particular in the effect of four-fermion operators of the third generation. 
These are, in the basis of \cite{Grzadkowski:2010es}, 
\begin{align}
\Delta \mathcal{L}_{\mathrm{SMEFT}}^{d=6}&=\frac{C_{tt}}{\Lambda^2}(\bar{t}_R \gamma_{\mu} t_R)(\bar{t}_R \gamma^{\mu} t_R)+\frac{C_{Qt}^{(1)}}{\Lambda^2}(\bar{Q}_L \gamma_{\mu} Q_L)(\bar{t}_R \gamma^{\mu} t_R)+\frac{C_{Qt}^{(8)}}{\Lambda^2}(\bar{Q}_L T^A\gamma_{\mu} Q_L)(\bar{t}_R T^A\gamma^{\mu} t_R) \nonumber\\ \label{eq:Lag}
&+\frac{C_{QQ}^{(1)}}{\Lambda^2}(\bar{Q}_L \gamma_{\mu} Q_L)(\bar{Q}_L \gamma^{\mu} Q_L)+\frac{C_{QQ}^{(3)}}{\Lambda^2}(\bar{Q}_L \sigma_a \gamma_{\mu} Q_L)(\bar{Q}_L  \sigma_a \gamma^{\mu} Q_L)\\ &+\left[\frac{C^{(1)}_{QtQb}}{\Lambda^2} (\bar{Q}_L t_R )i\sigma_2(\bar{Q}_L^{\rm T} b_R) +\frac{C^{(8)}_{QtQb}}{\Lambda^2} (\bar{Q}_L T^A t_R) i\sigma_2 (\bar{Q}_L^{\rm T} T^A b_R) + \hc \right]\,\nonumber
\\
&+ \frac{C_{bb}}{\Lambda^2}(\bar{b}_R \gamma_{\mu} b_R)(\bar{b}_R \gamma^{\mu} b_R) +\frac{C_{tb}^{(1)}}{\Lambda^2}(\bar{t}_R \gamma_{\mu} t_R)(\bar{b}_R \gamma^{\mu} b_R)+\frac{C_{tb}^{(8)}}{\Lambda^2}(\bar{t}_R T^A\gamma_{\mu} t_R)(\bar{b}_R T^A\gamma^{\mu} b_R) \nonumber \\
& +\frac{C_{Qb}^{(1)}}{\Lambda^2}(\bar{Q}_L \gamma_{\mu} Q_L)(\bar{b}_R \gamma^{\mu} b_R)+\frac{C_{Qb}^{(8)}}{\Lambda^2}(\bar{Q}_L T^A\gamma_{\mu} Q_L)(\bar{b}_R T^A\gamma^{\mu} b_R) \,,\nonumber
\end{align}
where we assume all Wilson coefficients to be real. In \eq{eq:Lag}, $Q_L$, $t_R$ and $b_R$ refer to the third family quark left-handed doublet and right-handed singlets, respectively; $\sigma_{a}$ are the Pauli matrices; $T^{A}$ are the $SU(3)_c$ generators and ${}^{\rm T}$ denotes transposition of the $SU(2)_L$ indices.

The largest effects in Higgs physics are typically expected to come from operators with the adequate chiral structure entering in top quark loops, as they will be proportional to the top quark mass/Yukawa coupling. 
Conversely, we expect a suppression of operators including bottom quarks with the bottom Yukawa coupling. 
As we will argue below, either because of their chirality or because they only enter in bottom loops, the operators with right-handed bottom quarks in the last two lines in \eq{eq:Lag} are expected to give only very small effects, and will be neglected. 
This is not the case for the operators  ${\cal O}_{QtQb}^{(1),(8)}$, which can have sizeable contributions to, e.g. Higgs to $b\bar{b}$ or gluon fusion rates, proportional to the top quark mass. 

We will later on also compare with possible effects of a trilinear Higgs self-coupling modification with respect to the SM. In the dimension-six SMEFT, the only operator that modifies the Higgs self-interactions without affecting the single-Higgs couplings at tree level is
\begin{equation}
\Delta \mathcal{L}_{\mathrm{SMEFT}}^{d=6}=\frac{C_{\phi}}{\Lambda^2}(\phi^\dagger \phi)^3,
\end{equation}
where $\phi$ stands for the usual $SU(2)_L$ scalar doublet, with $\phi=1/\sqrt{2}(0,v+h)^T$ in the unitary gauge.
Furthermore, for later use, we write down also the operators that modify the Higgs couplings to top and bottom quarks
\begin{equation}
\begin{split}
\Delta \mathcal{L}_{\mathrm{SMEFT}}^{d=6}&= \left(\frac{C_{t\phi}}{\Lambda^2}\phi^\dagger \phi ~\!\bar{Q}_L\tilde{\phi}~\!t_R+\frac{C_{b\phi}}{\Lambda^2}\phi^\dagger \phi ~\!\bar{Q}_L \phi~\! b_R+\hc\right)\,,
\end{split}
\end{equation}
with $\tilde{\phi}=i \sigma_2 \phi^*$.


\section{Contribution of four-quark operators to Higgs production and decay \label{sec:Higgs}}

In this section, we discuss the contribution of the third generation four-quark operators to various Higgs production mechanisms  and  Higgs decay channels.


\subsection{Higgs coupling to gluons and photons}

\begin{figure}[t!]
\begin{center}
\includegraphics[width=12cm]{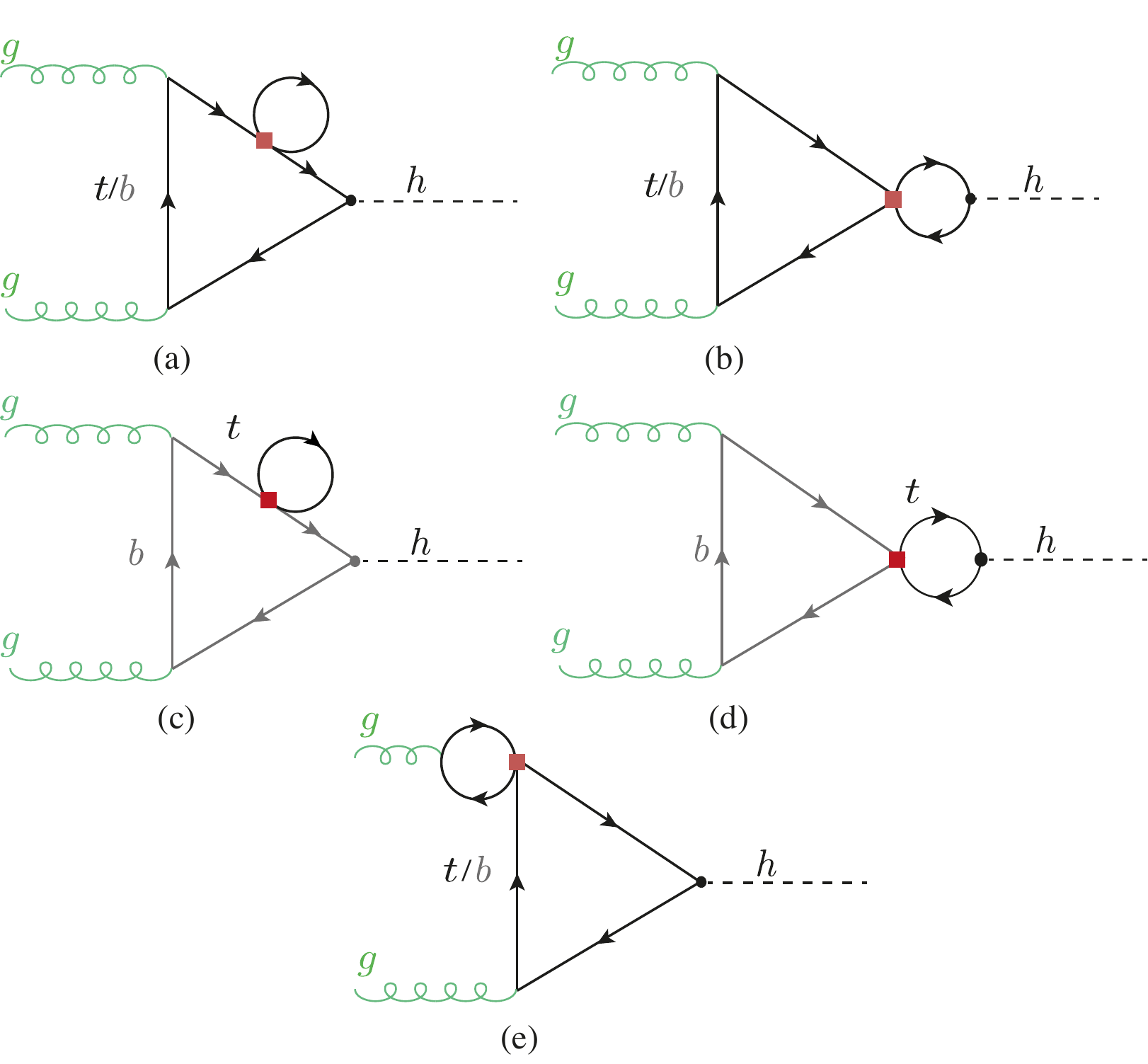}
\caption{Example Feynman diagrams for four-fermion-operator contributions to the Higgs production via gluon fusion. The red box indicates the four-fermion operator.\label{fig:ggh} }
\end{center}
\end{figure}

We start by discussing the calculation of the Higgs couplings to gluons and photons. The four-quark operators enter these couplings at the two-loop level. The diagrams are shown in \autoref{fig:ggh}. There are three classes of diagrams: (a) corrections to the top-quark propagator, (b) corrections to the Higgs Yukawa coupling and (c) corrections to the $t\bar{t}g$ and $t\bar{t}\gamma$ vertices. The latter turns out to be zero when the gluons or photons are on-shell.
The first and second type of corrections are left-right transitions hence the only contributions stem from the operators with Wilson coefficients $C_{Qt}^{(1),(8)}$, $C^{(1),(8)}_{QtQb}$ and $C_{Qb}^{(1),(8)}$.
As can be inferred from the diagrams in \autoref{fig:ggh} the result can be expressed as a product of one-loop integrals. We computed the diagrams in two independent calculations making use of different computer algebra tools such as \texttt{PackageX} \cite{Patel:2015tea}, \texttt{KIRA} \cite{Maierhoefer:2017hyi}, \texttt{Fire} \cite{Smirnov:2008iw}, \texttt{FeynRules} \cite{Alloul:2013bka} and \texttt{FeynArts} \cite{Hahn:2000kx}.\footnote{Note that the latter tool needed some manual adjustments to deal with four-fermion operators.}  
We cross-checked the Feynman rules with ref.~\cite{Dedes:2017zog}.
\\
For the renormalisation procedure we adopt a mixed on-shell (OS)-\MSbar -- scheme as proposed in \cite{Dawson:2018pyl}, in which we renormalise 
the quark masses OS and the Wilson coefficients of the dimension-six operators using  the \MSbar\;scheme. 
We hence renormalise the top/bottom mass as
\begin{equation}
m_{t/b}^{\text{OS}}=m_{t/b}^{(0)}-\delta m_{t/b},
\end{equation}
where the counterterms are given by
\begin{align}
\delta m_t =&\frac{1}{16 \pi^2} \frac{C_{Qt}^{(1)}+c_F C_{Qt}^{(8)}}{\Lambda^2}m_t^3\left[ \frac{2}{\bar{\epsilon}} +2 \log\left(\frac{\mu_R^2}{m_t^2}\right)+1\right] \\ &+ \frac{1}{16 \pi^2}  \frac{(2 N_c+1) C_{QtQb}^{(1)}+c_F C_{QtQb}^{(8)}}{\Lambda^2}  \left[ \frac{1}{\bar{\epsilon}} +  \log\left(\frac{\mu_R^2}{m_b^2}\right)+1 \right]  m_b^3\,, \nonumber \\
\delta m_b=&\frac{1}{16 \pi^2} \frac{(2 N_c+1)C_{QtQb}^{(1)}+c_F C_{QtQb}^{(8)}}{\Lambda^2}\left[ \frac{1}{\bar{\epsilon}} +\log\left( \frac{\mu_R^2}{m_t^2}\right)+1\right] m_t^3\,,
\end{align}
with $\bar{\epsilon}^{-1} = \epsilon^{-1}- \gamma_E +\log(4 \pi)$, in dimensional regularization with $d=4-2\epsilon$, $N_c=3$ the number of colours, and $c_F=(N_c^2-1)/(2N_c)=4/3$ the $SU(3)$ quadratic Casimir  in the fundamental representation. 
We note that, for the calculations of the physical processes in this paper, the difference between using the OS or the \MSbar\;definitions of the top and bottom masses in SMEFT results in changes that are formally of $\mathcal{O}(1/\Lambda^4)$.\footnote{In the \MSbar\;scheme the mass counterterms become
\begin{align}
\delta m_t^{\bar{\text{MS}}} =&\frac{1}{8 \pi^2} \frac{C_{Qt}^{(1)}+c_F C_{Qt}^{(8)}}{\Lambda^2}m_t^3\frac{1}{\bar{\epsilon}}+ \frac{1}{16 \pi^2}  \frac{(2 N_c+1) C_{QtQb}^{(1)}+c_F C_{QtQb}^{(8)}}{\Lambda^2}   \frac{1}{\bar{\epsilon}}  m_b^3\,,  \\
\delta m_b^{\bar{\text{MS}}}=&\frac{1}{16 \pi^2} \frac{(2 N_c+1)C_{QtQb}^{(1)}+c_F C_{QtQb}^{(8)}}{\Lambda^2}\frac{1}{\bar{\epsilon}} m_t^3\,.
\end{align}
}
We note though that using a SM running \MSbar\;bottom mass instead of an OS one makes a relevant difference in the numerical results.
In the results presented below we will use the OS bottom mass as an input.

The coefficients of the dimension-six operators 
are renormalised in the \MSbar \;scheme.
At one-loop level the only operators entering the Higgs to gluon or photon rates that mix with the four-quark operators are the ones that modify the top or bottom Yukawa couplings: ${\cal O}_{t\phi}$ and ${\cal O}_{b\phi}$, respectively. The coefficients of these operators are renormalized according to
\begin{equation}
C^{\bar{\text{MS}}}_{t\phi/b\phi}=C^{(0)}_{t\phi/b\phi}+\delta C_{t\phi/b\phi}\quad\quad \text{   with   }\quad\quad \delta C_{t\phi/b\phi} = -\frac{1}{2\bar{\epsilon}}\frac{1}{16 \pi^2} \gamma^{j}_{t\phi/b\phi} C_j.
\end{equation}
The only four-quark Wilson coefficients contributing to $\gamma_{t\phi/b\phi}$ are the ones from ${\cal O}_{Qt}^{(1),(8)}$,  ${\cal O}_{QtQb}^{(1),(8)}$ and ${\cal O}_{Qb}^{(1),(8)}$. 
The explicit expressions for the relevant one-loop anomalous dimension can be obtained from ref.~\cite{Jenkins:2013zja,Jenkins:2013wua}. 
The Wilson coefficients $C_{t\phi/b\phi}$ modify the Higgs couplings to top quarks/bottom quarks as follows
\begin{equation}
g_{ht\bar{t}/hb\bar{b}}=\frac{m_{t/b}}{v}-\frac{v^2}{\Lambda^2}\frac{C_{t\phi/b\phi}}{\sqrt{2}}\,.
\end{equation}
Hence, a modification of the Higgs couplings to bottom and top quarks is generated by operator mixing, even if $C_{t\phi/b\phi}$ are zero at $\Lambda$.
\par
The modification of the Higgs production rate in gluon fusion (ggF) can be written as
\begin{equation}
\frac{\sigma_{ggF}}{\sigma_{ggF}^{\SM}}= 1+ \frac{ \sum_{i=t,b} 2 \Re(F_{\LO}^i F^*_{\NLO})}{\left| F_{\LO}^t+F_{\LO}^b  \right|^2}, \label{eq:production}
\end{equation}
with 
\begin{equation}
F_{\LO}^i=-\frac{8m_i^2}{m_h^2}\left[1-\frac{1}{4}\log^2(x_i)\left(1-\frac{4m_i^2}{m_h^2}\right)\right],
\end{equation}
where $m_h$ is the Higgs mass,
and
\begin{equation}
\begin{split}
F_{\NLO}=&\frac{ 1}{4\pi^2  \Lambda^2}(C_{Qt}^{(1)}+c_FC_{Qt}^{(8)})F_{\LO}^t \Bigg[ 2 m_t^2  +\frac{1}{4} (m_h^2-4 m_t^2) \left( 3 +2 \sqrt{1-\frac{4 m_t^2}{m_h^2}} \log(x_t) \right)   \\ & 
+\frac{1}{2} (m_h^2-4 m_t^2) \log\left(\frac{\mu_R^2}{m_t^2}\right)\Bigg] \\ & + 
\frac{1}{32 \pi^2 \Lambda^2} ((2N_c+1)C_{QtQb}^{(1)}+c_FC_{QtQb}^{(8)}) \Bigg[ F_{\LO}^b \frac{m_t}{m_b}\Bigg( 4 m_t^2-2 m_h^2  \\ &  - (m_h^2-4 m_t^2)\sqrt{1-\frac{4 m_t^2}{m_h^2}} \log(x_t)-(m_h^2-4 m_t^2)\log\left(\frac{\mu_R^2}{m_t^2}\right)\Bigg) +(t\leftrightarrow b)\Bigg]  \,. \label{eq:FNLO}
\end{split}
\end{equation}
Only top quark loops contribute to the parts proportional to $C_{Qt}^{(1),(8)}$. 
We have neglected the contributions of the operators with Wilson coefficient $C_{Qb}^{(1),(8)}$ as they would lead only to subleading contributions proportional to $m_b^3$.
The variable $x_i$ for a loop particle with mass $m_i$ is given by
\begin{equation}
x_i=\frac{-1+\sqrt{1-\frac{4 m_i^2}{m_h^2}}}{1+\sqrt{1-\frac{4 m_i^2}{m_h^2}}}\,. \label{eq:xvariable}
\end{equation} 
In analogy to \eq{eq:production}, we can write the modified decay rates of the Higgs boson to gluons as
\begin{equation}
\frac{\Gamma_{h\to gg}}{\Gamma_{h\to gg}^{\SM}}= 1+ \frac{ \sum_{i=t,b} 2 \Re(F_{\LO}^i F^*_{\NLO})}{ |F_{\LO}^t+F_{\LO}^b|^2} 
\end{equation}
and 
\begin{equation}
\frac{\Gamma_{h\to \gamma\gamma}}{\Gamma_{h\to \gamma\gamma}^{\SM}}= 1+ \frac{2 \Re(F_{\LO, \gamma} F^*_{\NLO,\gamma})}{  |F_{\LO, \gamma}|^2} .
\end{equation}
In the latter 
\begin{equation}
F_{\LO, \gamma}= N_C\,Q_t^2 F_{\LO}^t+ N_C\,Q_b^2 F_{\LO}^b+F_{\LO}^W+ F^G_{LO} ,
\end{equation}
and $F_{\NLO, \gamma}$ is obtained from $F_{\NLO}$ by replacing the LO form factor that appears inside of it by  $ F_{\LO}^i \to N_c \,Q_i^2 F_{\LO}^i$,
with the charges $Q_t=2/3$ and $Q_b=-1/3$. The $W$ boson contribution is given by
\begin{equation}
F_{\LO}^W= 2 \left(1+6 \frac{m_W^2}{m_h^2}\right)-6 \frac{m_W^2}{  m_h^2} \left(1-2  \frac{m_W^2}{m_h^2}\right) \log^2(x_W),
\end{equation}
with $m_W$ the $W$ mass, and the Goldstone contribution is
\begin{equation}
	F_{\LO}^G=4\frac{m_W^2}{m_h^2} \left( 1+ \frac{m_W^2}{m_h^2} \,\log^2(x_W) \right)\,.
\end{equation}
\par

The formulae presented above are valid under the assumption that, at the electroweak scale, the four-quark operators are the only new physics contributions in the dimension-six effective Lagrangian. If, on the other hand, one assumes that the four-quark operators are defined at some higher scale $\Lambda$, e.g.~after matching with an specific ultraviolet (UV) model, further (logarithmic) contributions appear during the running to low energies, as a result of the mixing between these four-fermion interactions and those operators that would modify the processes at LO. 
Those effects can be included via the renormalisation group equation (RGE) for the operators with Wilson coefficient $C_{t\phi}$  and $C_{b\phi}$  \cite{Jenkins:2013zja, Jenkins:2013wua}, that lead approximatively to 
\begin{equation}
\begin{split}
C_{t\phi}(\mu_R)-C_{t\phi}(\Lambda)= &\frac{1}{16 \pi^2 v^2} \left[-2  y_t (m_h^2  -4 m_t^2) (C_{Qt}^{(1)}+c_F C_{Qt}^{(8)} )\log\left( \frac{\mu_R^2}{\Lambda^2}\right) \right.\\
& \left.+ \frac{y_b}{2} (m_h^2-4 m_b^2)\left(  (2N_c+1)  C_{QtQb}^{(1)}+   c_F C_{QtQb}^{(8)}\right)\log\left( \frac{\mu_R^2}{\Lambda^2}\right)\right] \label{eq:runningCuH}
\end{split}
\end{equation}
and
\begin{equation}
\begin{split}
C_{b\phi}(\mu_R)-C_{b\phi}(\Lambda)= \frac{y_t}{32 \pi^2 v^2} \left[  (m_h^2-4 m_t^2)\left(  (2N_c+1)  C_{QtQb}^{(1)}+   c_F C_{QtQb}^{(8)}\right)\log\left( \frac{\mu_R^2}{\Lambda^2}\right)\right]\,, \label{eq:runningCdH}
\end{split}
\end{equation}
where $y_{t/b}=\sqrt{2} m_{t/b}/v$, and we have neglected contributions from $C_{Qb}^{(1),(8)}$ in \eq{eq:runningCdH}, as they are proportional to $y_b$, and thus lead to corrections to the rates without any $m_t$ enhancement.
Note that the combinations of Wilson coefficients  appearing in \eqs{eq:runningCuH}{eq:runningCdH} are the same as in $F_{\rm NLO}$ in \eq{eq:FNLO}.
Effectively, we can then obtain the result under the assumption that the four-fermion operators are the only non-zero ones at the high scale by replacing in \eq{eq:FNLO} $\mu_R \to \Lambda$, noting that we have renormalised the top and bottom quark mass in the OS scheme.

\subsection{Higgs decay to bottom quarks}
   \begin{figure}[t!]
   	\vspace{-.5 cm}
	\centering
	\includegraphics[scale=0.85]{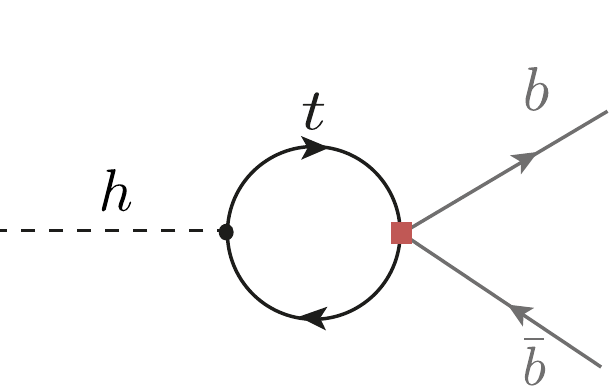}
	\caption{Feynman diagram contributing to the NLO   $h \to b \bar b$ process. }
	\label{hbb}
\end{figure}
The dominant four-fermion contributions to the decay channel $h \to b\bar b$ come from the operators $\mathcal O_{QtQb}^{(1),(8)}$. The corresponding diagram at NLO is shown in~\autoref{hbb}. 
Adopting the same renormalisation procedure as outlined in the previous subsection, we obtain the following expression for the correction to 
the $h \to b\bar b$ decay rate in the presence of ${\cal O}_{QtQb}^{(1),(8)}$,
\begin{equation}
\begin{split}
\frac{\Gamma_{h\to b\bar{b}}}{\Gamma_{h\to b\bar{b}}^{\SM}}=& 1+ \frac{1}{16\pi^2}\frac{m_t}{m_b}(m_h^2-4m_t^2)\frac{(2 N_c+1) C_{QtQb}^{(1)}+c_F C_{QtQb}^{(8)}}{\Lambda^2} \\ & \times\left[ 2+\sqrt{1-\frac{4 m_t^2}{m_h^2}}\log(x_t)-\log\left(\frac{m_t^2}{\mu_R^2}\right) \right] \,,
\label{hbbnlo}
\end{split}
\end{equation}
which carries an enhancement factor of $m_t/m_b$ and is hence expected to be rather large.
Again, we have neglected subdominant contributions suppressed by the bottom mass from the operators $\mathcal{O}_{Qb}^{(1),(8)}$. 
Including the leading logarithmic running of $C_{b\phi}$ of \eq{eq:runningCdH} from the high scale $\Lambda$ to the electroweak scale is achieved by setting in \eq{hbbnlo} $\mu_R\to \Lambda$.
The expression in \eq{hbbnlo} agrees with the results obtained from a full calculation of the NLO effects in the dimension-six SMEFT, first computed in~\cite{Gauld:2015lmb}. 

This closes the discussion of the main effects that the third-generation four-quark operators can have in the different Higgs decay widths.\footnote{Four-fermion operators also affect the $h\to Z\gamma$ partial width. However, as in the diphoton case, the effect is expected to be small due to the dominance of the $W$ boson loop. Because of this, and given the smallness of the $h\to Z\gamma$ branching ratio and the relatively low precision expected in this channel at the LHC, we neglect the effects of four-fermion interactions in this decay.} Note also that these modifications of the Higgs decay rate to photons, gluons and, especially, bottom quarks, affect all the branching ratios (BRs) due to the modification of the Higgs total width, and therefore have an observable effect in all Higgs processes measured at the LHC.


\subsection{Associated production of a Higgs boson with top quarks}

   \begin{figure}[t!]
	\centering
	\includegraphics[scale=0.85]{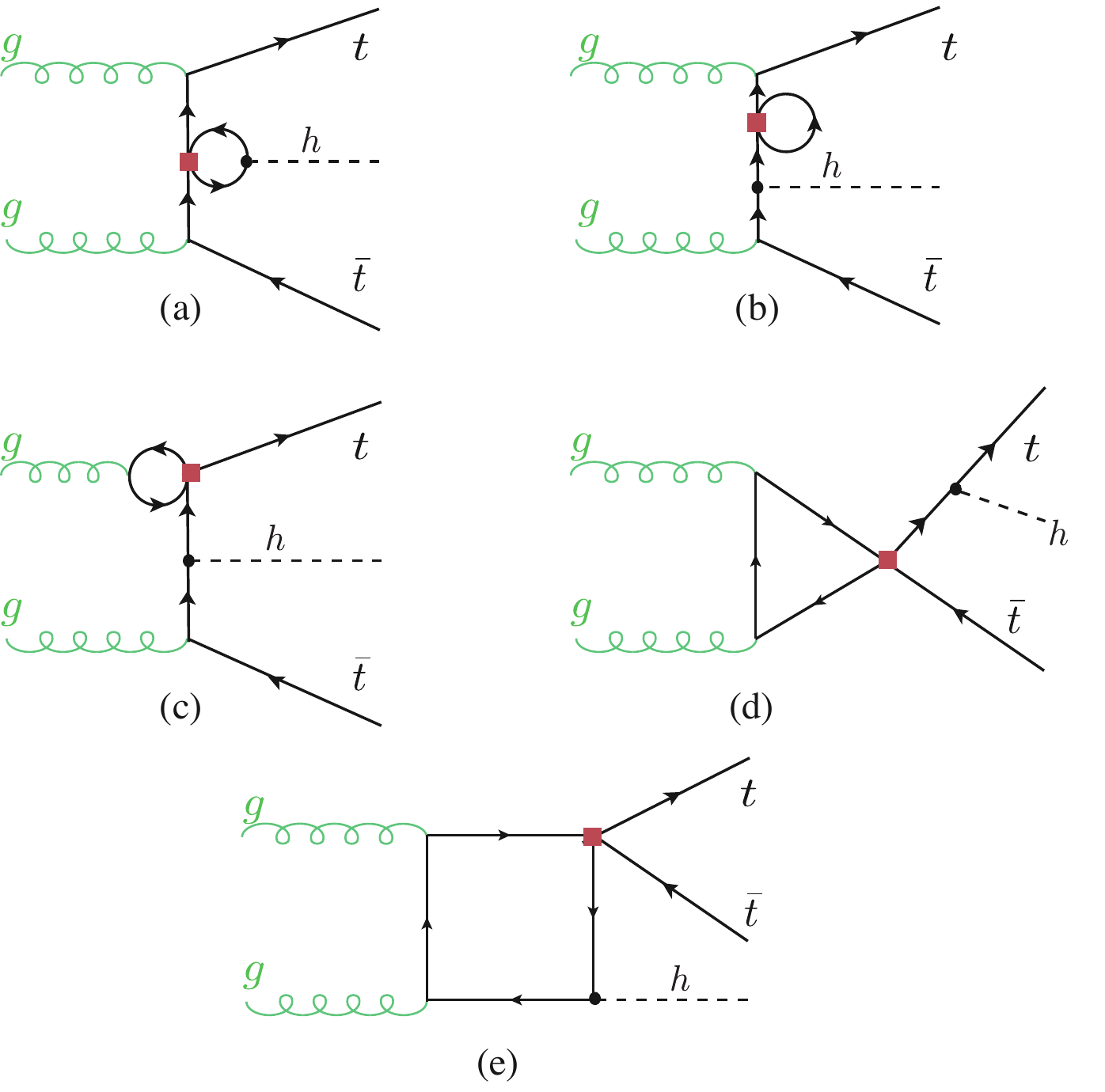}
	\caption{Example Feynman diagrams including the four-fermion loop contributions to the $ gg \to t\bar{t} h$ subprocess. }
	\label{ggtth}
\end{figure}
   \begin{figure}[ht!]
	\centering
	\includegraphics[scale=0.85]{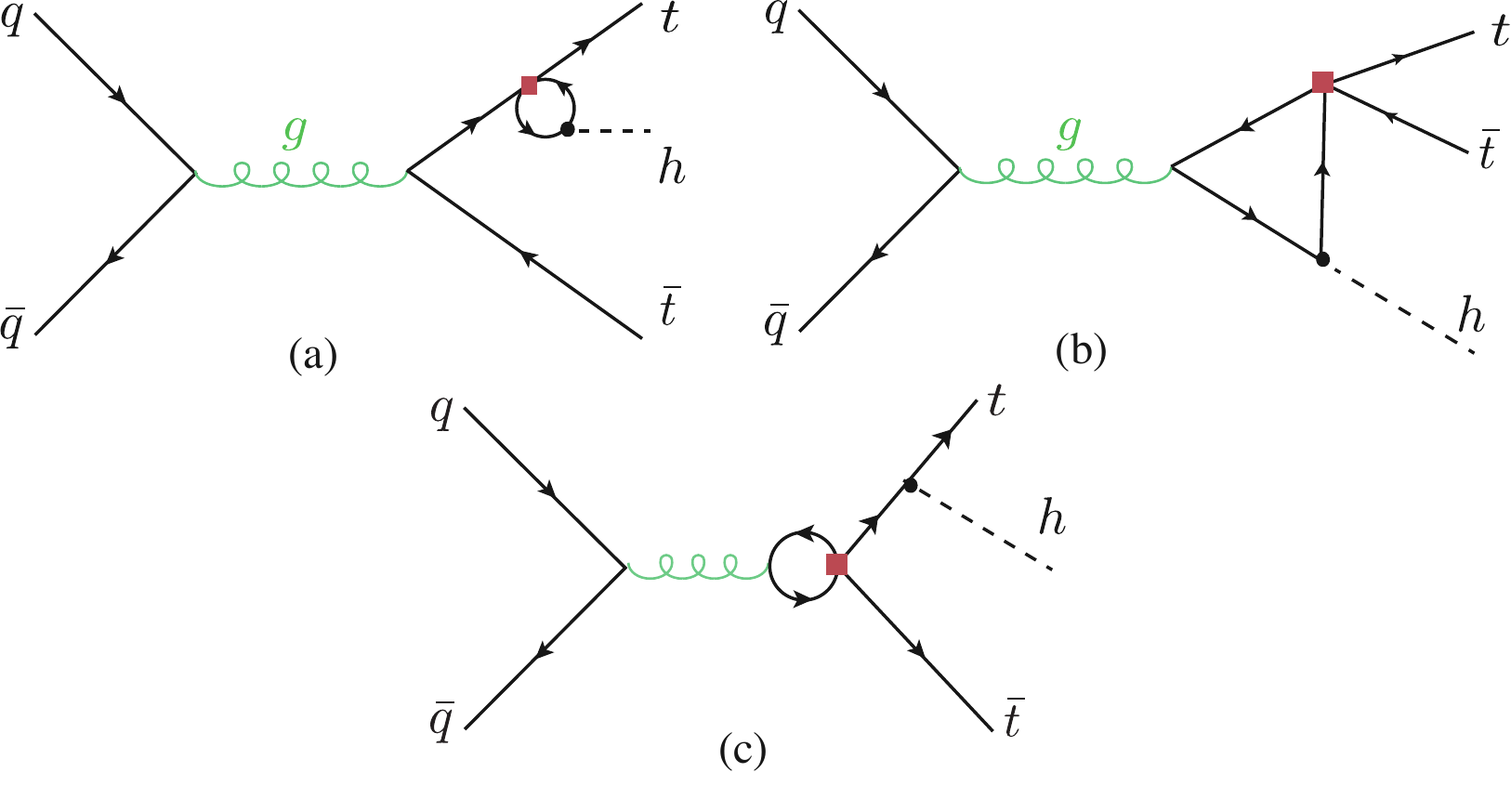}
	\caption{Example Feynman diagrams including the four-fermion loop contributions to the  $ q \bar{q} \to t\bar{t} h$ subprocess.  }
	\label{qqtth}
\end{figure}

 The $ t\bar{t} h$ process receives NLO corrections from four-quark operators from a large number of diagrams. The process can be initiated either by gluons, see \autoref{ggtth} for a sample of the corresponding diagrams, or by a quark anti-quark pair, see \autoref{qqtth}.  The triangle and box topologies (shown as (d) and (e) in \autoref{ggtth} and as (b) in \autoref{qqtth}) are finite. 
 We have computed the leading NLO contributions for both types of processes, via the interference of the four-quark loops with the LO QCD amplitudes. 
 For the computation of the quark-initiated contributions we adopt a four-flavour scheme  We note that within a five-flavour scheme operators containing both bottom and top quarks lead to a LO contribution from a direct contact diagram. Nevertheless, this gives an overall negligible correction as the $ b \bar b $ initiated $t\bar t h$ process is suppressed by the small bottom parton distribution functions. The effect of changing the flavour scheme results in an uncertainty of $1-2\%$.

 The NLO effects were obtained via an analytic computation\footnote{The {\tt FORTRAN} code containing this analytical calculation can be provided on request.}, based on the reduction of one-loop amplitudes via the method developed by G. Ossola, C.G. Papadopoulos and R. Pittau~(OPP reduction)~\cite{Ossola:2006us}.
 The OPP reduction was done using the \texttt{CutTools} programme~\cite{Ossola:2007ax}.
It reduces the one-loop amplitude into 1,2,3 and 4-point loop functions in four dimensions, keeping spurious terms from the $\bar{\epsilon}$ part of the amplitude. To correct for such terms, one needs to compute the divergent UV counterterm as well as a finite rational terms, denoted $R_{2}$ as in ref.~\cite{Ossola:2008xq}.\footnote{Another rational term ~$R_1$ appears due to the mismatch between the four and $d$ dimensional amplitudes, but this is computed automatically in \texttt{CutTools}. } The amplitudes were generated in the same way as for gluon fusion. The UV and $R_2$ counterterms, that need to be supplemented to \texttt{CutTools}, were computed manually following the method detailed in~\cite{Ossola:2008xq}.
The UV counterterms are the same as for gluon fusion, in addition to a new one  that is needed to be introduced to renormalise diagrams of type (c) in \autoref{ggtth} and \autoref{qqtth}. This is due to the operator mixing of light -- heavy four-quark operators with heavy four-quark operators. Effectively, this leads to a counterterm 
  \begin{equation}
	\includegraphics[width=0.13\linewidth]{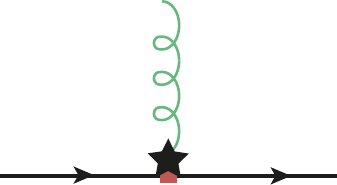} =\frac{ig_s}{12 \pi^2\Lambda^2} T^{A}_{ij} p_g^2 \gamma^\mu \left( C_{tt} P_R+\left(C_{QQ}^{(1)} + C_{QQ}^{(3)}\right) P_L +\frac{ C_{Qt}^{(8)}}{4} \right) \left( \frac{1}{\bar{\epsilon}}-1 \right) . \label{eq:R2CT}
 \end{equation}

\par
 The results of our calculation were cross-checked using \texttt{Madgraph\_aMCNLO} \cite{Alwall:2014hca} (version  3.1.0) using the \texttt{SMEFTatNLO v1.0.2} model~\cite{Degrande:2020evl}. To match both calculations, we set {\tt MadLoop} to filter out the NLO QCD corrections, whereas other contributions such as those in diagrams (a) and (b) in 
\autoref{ggtth}, or (a) and (c) in \autoref{qqtth}, needed to be included, as they are filtered out by the default {\tt MadLoop} settings.

The results of our calculation reveal that associated Higgs production with top quarks receives significant finite NLO corrections from the singlet and octet operators $\mathcal{O}_{Qt}^{(1),(8)}$, while the contributions from other operators, e.g.~the singlet and triplet left-handed operators, $\mathcal{O}_{QQ}^{(1),(3)}$, or the right-handed four-top operator, $\mathcal{O}_{tt}$, are small. As will also be seen in the explicit numerical results presented in the next section, while for Higgs production via ggF or the decay into gluons only certain combinations of singlet/octet operators entered, leading to a degeneracy, this is not the case for $t\bar{t}h$ production, where the gluons no longer need to combine to a colour singlet state. 
The degeneracy between the singlet and octet operators is mainly broken by the contributions from the triangle diagrams, where, for instance, the difference between the contributions of $\mathcal{O}_{Qt}^{(1)}$ and $\mathcal{O}_{Qt}^{(8)}$ does not follow the same colour structure as other diagrams. 

As we saw in the previous sections, the operators $\mathcal{O}_{QtQb}^{(1),(8)}$ are expected to have a sizeable effect in ggF and, in particular, $h\to b\bar{b}$. This is not the case for $t\bar{t}h$, where we explicitly computed their contributions and found them to be negligible, as they are suppressed by the $b$-quark mass.  
Similarly, other ``mixed'' bottom-top operators are expected
to give very suppressed contributions compared to those from four-top operators. Therefore we neglected their effects in our calculation.\footnote{
In this regard, since the singlet and octet operators $\mathcal{O}_{QtQb}^{(1),(8)}$ are not implemented in the current version of \texttt{SMEFTatNLO}, or in any other loop-capable \texttt{UFO} model available, we have modified the \texttt{SMEFTatNLO} model to include these operators, by including their Feynman rules and computing the UV and $R_2$ counterterms needed for the $t\bar t h$ calculation. 
The other ``mixed'' bottom-top operators are also currently not included in \texttt{SMEFTatNLO}. A computation of their contributions, while being beyond the scope of this paper, would require a similar strategy as for the $\mathcal{O}_{QtQb}^{(1),(8)}$ operator.}

Again, to connect with specific models that may generate the four-quark operators at the new physics scale $\Lambda$, one needs to consider the contributions that come from the running from $\Lambda$ to low energies, and that mix these operators with those entering in $t\bar{t}h$ at the LO level. 
For the gluon-initiated subprocess the relevant contributions are from the running of $C_{t\phi}$ in \eq{eq:runningCuH}, while for the quark-initiated subprocess we also need to account for the mixing of the third generation four-fermion operators with the ones connecting the third generation with the first two generations. The corresponding corrections can be obtained from the RGEs in refs. \cite{Jenkins:2013zja,Jenkins:2013wua,Alonso:2013hga}. As in the case of the finite pieces, the logarithmic corrections from top-bottom operators were found to be very small and are neglected in what follows.


\subsection{Results}

Here we provide semi-analytical expressions for the results of our NLO calculations including the effects of the third generation four-quark operators. 
These NLO contributions to the single-Higgs rates, as a function of the four-heavy-quark Wilson coefficients, are denoted by
 \begin{equation}
\delta R(C_i) = R/R^{\SM} -1,
\label{eq:deltaR}
\end{equation}
where $R$ stands generically for a given partial width $\Gamma$ or cross section $\sigma$. They are summarised in~\autoref{table:res4top}. 
 The results for $\delta R(C_i)$ in that table concern only the linear contributions in $\Lambda^{-2}$. 
 They have been written as
 \begin{equation}
 \delta R(C_i)= \frac{C_i}{\Lambda^2}\left(\delta R_{C_i}^{fin}+ \delta R_{C_i}^{log} \log\left(\frac{\mu_R^2}{\Lambda^2}\right)\right)\,,
 \label{eq:deltar}
 \end{equation}
 where we have separated the contributions in two parts: the first, parameterised by $\delta R_{C_i}^{fin}$, concerns the finite part of the NLO correction taken at the typical scale $\mu_R$ of the process; the second, parameterised by $\delta R_{C_i}^{log}$, is the logarithmic contribution, obtained by solving the RGE of the dimension-six Wilson coefficients from the high scale $\Lambda$ to the low scale $\mu_R$, using the leading log approximation.
Both the finite part dependence $\delta R_{C_i}^{fin}$ of these corrections on the Wilson coefficient as well as the part proportional to the logarithm $\delta R_{C_i}^{log}$  are reported in~\autoref{table:res4top}.
 Our results can be improved by replacing the part proportional to the coefficients  $\delta R^{log}_{C_i}$ by the solution of the coupled system of RGEs. 

For $ \Lambda =1$ TeV, and depending on the renormalisation scale of the process, the value of the logarithm in \eq{eq:deltar} ranges between $\sim[-5.5,-2.9]$. With these numerical values in mind and by looking at $\delta R_{C_i}^{log}$ in~\autoref{table:res4top}, we see that the finite part of the NLO calculation, i.e. $\delta R_{C_i}^{fin}$, is usually of the same order  of magnitude than the leading-log part. The clear exceptions are the $C_{Qt}^{(1),(8)}$ contributions to $t\bar{t}h$, where the finite pieces dominate, and the $C_{QtQb}^{(1),(8)}$ contributions to the $h\to b\bar{b}$, where the logarithmic contributions are the leading ones. This underlines the importance of considering the full NLO computation in the determination of the Wilson coefficients for $C_{Qt}^{(1),(8)}$, whereas for $C_{QtQb}^{(1),(8)}$, where the limits are mainly driven by $h\to b\bar{b}$, they turn out to play a less important role.
 
\begin{table}[t!]
	\centering
	\small{
		\begin{tabular}{c||cccc}
			\toprule
			{ \normalsize Operator} &  { \normalsize Process }& { \normalsize $\mu_R$} & { \normalsize$ \delta R_{C_i}^{fin}\; [\text{TeV}^2]$} &{ \normalsize$ \delta R_{C_i}^{log}\; [\text{TeV}^2] $} \\
			\midrule
            \multirow{5}{*}{ { \normalsize$\mathcal{O}_{Qt}^{(1)}$}}  &  ggF& $\frac{m_h}{ 2}$&$\phantom{+}9.91\cdot 10^{-3}$&$\phantom{+}2.76\cdot 10^{-3}$\\     
                                                                    &  $h \to gg$& \mr{$m_h$}&$\phantom{+}6.08\cdot 10^{-3}$&$\phantom{+}2.76\cdot 10^{-3}$\\
            	                                                   &  $h \to \gamma \gamma$& &$-1.76\cdot 10^{-3}$ &$-0.80\cdot 10^{-3}$ \\
            	                                                   	&  $t\bar t h$ {\color{Mahogany}  13 TeV }&\mr{ $m_t+\frac{m_h}{ 2}$}&$-4.20\cdot 10^{-1} $&$-2.78\cdot 10^{-3}$\\	    
            	                                                   	&   $t\bar t h$  {\color{Mahogany}  14 TeV }& &$-4.30\cdot 10^{-1} $&  $-2.78\cdot 10^{-3}$\\	
            	                                                   	\midrule
          \multirow{5}{*}{ { \normalsize$\mathcal{O}_{Qt}^{(8)}$} } & ggF& {$\frac{m_h}{ 2}$}&$\phantom{+}1.32\cdot 10^{-2}$&$\phantom{+}3.68\cdot 10^{-3}$\\    
                                                                   &  $h \to gg$& \mr{$m_h$}&$\phantom{+}8.11\cdot 10^{-3}$&$\phantom{+}3.68\cdot 10^{-3}$\\
            	                                                   	&  $h \to \gamma \gamma$& &$-2.09\cdot 10^{-3}$&$-1.07\cdot 10^{-3}$\\
            	                                                   	&  $t\bar t h$ {\color{Mahogany}  13 TeV }& \mr{$m_t+\frac{m_h}{ 2}$}&$\phantom{+}6.81\cdot 10^{-2}$ &$-2.40\cdot 10^{-3}$\\	    
            	                                                   	&   $t\bar t h$  {\color{Mahogany}  14 TeV }& & $\phantom{+}7.29\cdot 10^{-2}$&  $-2.48\cdot 10^{-3}$\\	              
                       	                                                   	\midrule
           \multirow{4}{*}{ { \normalsize$\mathcal{O}_{QtQb}^{(1)}$} } & ggF& ${m_h\over 2}$&$\phantom{+}2.84\cdot 10^{-2}$&$\phantom{+}9.21\cdot 10^{-3}$\\   
            &  $h \to gg$& \multirow{3}{*}{$m_h$}&$\phantom{+}1.57\cdot 10^{-2}$&$\phantom{+}9.21\cdot 10^{-3}$\\
           &  $h \to \gamma \gamma$& &$-1.30\cdot 10^{-3}$&$-0.78\cdot 10^{-3}$\\
           &  $h \to b \bar b$& &$\phantom{+}9.25\cdot 10^{-2}$&$\phantom{+}1.68\cdot 10^{-1}$\\
			\midrule
			 \multirow{4}{*}{{ \normalsize$\mathcal{O}_{QtQb}^{(8)}$}}  & ggF& {$\frac{m_h}{ 2}$}&$\phantom{+}5.41\cdot 10^{-3}$&$\phantom{+}1.76\cdot 10^{-3}$\\      
			 & $h \to gg$& \multirow{3}{*}{$m_h$}&$\phantom{+}2.98\cdot 10^{-3}$&$\phantom{+}1.76\cdot 10^{-3}$\\
			&  $h \to \gamma \gamma$& &$-0.25\cdot 10^{-3}$& $-0.15\cdot 10^{-3}$\\
			&  $h \to b \bar b$& &$\phantom{+}1.76\cdot 10^{-2}$&$\phantom{+}3.20\cdot 10^{-2}$\\
			\midrule	    	 
			 \multirow{2}{*}{{ \normalsize$\mathcal{O}_{QQ}^{(1)}$}  }
			 	&  $t\bar t h$ {\color{Mahogany}  13 TeV }& \mr{$m_t+\frac{m_h}{ 2}$}&  {$\phantom{+}1.75\cdot 10^{-3}$} &$\phantom{+}1.84\cdot 10^{-3}$\\	    
			 &   $t\bar t h$  {\color{Mahogany}  14 TeV }& & $\phantom{+}1.65\cdot 10^{-3}$& $\phantom{+}1.76\cdot 10^{-3}$\\          
			 \midrule	    	 
			 \multirow{2}{*}{{ \normalsize$\mathcal{O}_{QQ}^{(3)}$}  }
			 &  $t\bar t h$ {\color{Mahogany}  13 TeV }& \mr{$m_t+\frac{m_h}{ 2}$}&  $\phantom{+}1.32\cdot 10^{-2}$ & $\phantom{+}5.48\cdot 10^{-3}$\\	    
			 &   $t\bar t h$  {\color{Mahogany}  14 TeV }& & $\phantom{+}1.24\cdot 10^{-2}$& $\phantom{+}5.30\cdot 10^{-3}$\\        
			  \midrule	    	 
			 \multirow{2}{*}{{ \normalsize$\mathcal{O}_{tt}$}  }
			 &  $t\bar t h$ {\color{Mahogany}  13 TeV }& \mr{$m_t+\frac{m_h}{ 2}$}&  $\phantom{+}4.60\cdot 10^{-3}$ &$\phantom{+}1.82\cdot 10^{-3}$\\	    
			 &   $t\bar t h$  {\color{Mahogany}  14 TeV }& & $\phantom{+}4.57\cdot 10^{-3}$& $\phantom{+}1.74\cdot 10^{-3}$\\                                           	
			\bottomrule
		\end{tabular}
	}
	\caption{The NLO corrections to single Higgs rates from the four heavy-quark SMEFT operators of this study. We have separated the contributions into the finite piece,~$ \delta R_{C_i}^{fin}$, and the leading log running of the Wilson coefficients,~$ \delta R_{C_i}^{log}$, see~\eq{eq:deltar}. The $ \delta R_{C_i}^{fin, log}$ terms for the ${\cal O}_{QtQb}^{(1),(8)}$ operators for the $t\bar{t}h$ process are of $\mathcal{O}(10^{-5}-10^{-6})$ TeV$^{-2}$ and are omitted in the table. }
\label{table:res4top}
\end{table}

The numerical values were obtained using as input parameters
 \begin{equation}
 \begin{split}
& G_F=1.166378 \cdot 10^{-5} \text{ GeV}^{-2}\,,  \; m_W=80.379\text{ GeV}\,, \;m_Z=91.1876\text{ GeV}\,, \\ &  m_t^{\text{OS}}=172.5 \text{ GeV}\,, \; m_b^{\text{OS}}=4.7\text{ GeV}\,,  \;m_h=125.1\text{ GeV}\,,
 \end{split}
 \end{equation}
where the OS bottom quark mass is taken from~\texttt{RunDec}~\cite{Chetyrkin:2000yt}, and the rest of the parameters from the particle data group~\cite{Zyla:2020zbs}. We have used  the NNPDF23 parton distribution functions set at NLO \cite{Ball:2012cx}. 

Looking at the results, first we note that the operators ${\cal O}_{QQ}^{(1),(3)} $ and ${\cal O}_{tt}$ only contribute to $t\bar t h$ production. 
In this regard, however, it must be noted that the uncertainties due to missing higher order corrections and the PDF+$\alpha_s$ uncertainty
for the $t\bar t h$ process are at the several percent level, $\sigma_{\rm tth, 13 TeV}^{\rm SM} =0.506^{+6.9\%}_{-10\%}~\!{\rm pb}$ \cite{LHCHiggsCrossSectionWorkingGroup:2016ypw}. This is larger than the typical finite effects of $C_{QQ}^{(1),(3)} $ and $C_{tt}$ for ${\cal O}(1)$ coefficients and $\Lambda \sim 1$ TeV. Therefore, all Higgs rates are expected to be relatively insensitive to these interactions 
unless rather large values of the Wilson coefficients are allowed.  
Secondly, from the analytic results, we observe that in the NLO corrections to Higgs decay rates and gluon fusion, the Wilson coefficients $C_{QtQb}^{(1)} ,  C_{QtQb}^{(8)}$ always appear in a linear combination identical to the one seen in the RGE of the Wilson coefficients~$C_{t\phi}$ and~$C_{b\phi}$, i.e. 
\begin{equation}
	   C_{QtQb}^+= (2N_c+1 )C_{QtQb}^{(1)} + c_F   C_{QtQb}^{(8)}.
	   \label{eq:CQtQbplus}
\end{equation}
In fact, all single-Higgs rates are mostly sensitive to the linear combination in \eq{eq:CQtQbplus}.
This is because, even though the $\mathcal{O}_{QtQb}^{(1),(8)}$ operators also enter in diagrams contributing to the $t\bar t h$ process, the corresponding finite corrections are suppressed by the bottom quark mass and therefore very small. (For these operators, the results for $\delta R_{C_i}^{fin,log}$ are of $O(10^{-5}-10^{-6})$ TeV$^{-2}$, and were omitted in~\autoref{table:res4top}.) This suppression is also expected for other ``mixed" top-bottom operators,
which would contribute to $t\bar t h$ via bottom-quark loops and hence would be strongly suppressed, justifying that we did not consider them here. 
In summary, apart from $\mathcal{O}_{Qt}^{(1),(8)}$, all the other third-generation four-quark operators produce only small contributions to the $t\bar{t}h$ process.

\section{Fit to Higgs observables \label{sec:fit}}

\par In this section we will show the results of a fit to Higgs observables of the four-quark operators of the third generation and the operator that modifies the Higgs potential and hence the Higgs self-coupling. In ref.~\cite{Gorbahn:2016uoy, Degrassi:2016wml, Bizon:2016wgr, Maltoni:2017ims, Degrassi:2021uik} it was proposed to extract the trilinear Higgs self-coupling via its loop effects in single-Higgs measurements. 
Within the assumptions of the SMEFT, a model-independent determination of the triple Higgs self-interaction, $\lambda_3$, should be considered within a global analysis considering all effective interactions that enter up to the same order in perturbation theory as $\lambda_3$.
In particular, apart from the trilinear Higgs self-coupling modification, such a study must include those operators that enter at LO in Higgs production and decay~\cite{DiVita:2017eyz}. Furthermore, the sensitivity to the Higgs self-coupling modifications can also be diminished by other operators entering as the trilinear Higgs self-coupling via loop effects, if those operators are not yet strongly constrained experimentally by other processes. Such is the case for some of the four-quark operators considered in this paper. In order to show this, we have performed a fit to Higgs data of the operator $\mathcal{O}_{\phi}$ and the four-fermion operators considered in this study. A full global fit including all new physics effects would require the combination of Higgs data with that from other processes and is beyond the scope of this paper. 

\subsection{Fit methodology}

For each experimentally observed channel with a signal strength~$\mu_{\mathrm{Exp}}\equiv \sigma_{\mathrm{Obs}}/\sigma_\SM$, one can build a theoretical prediction for this signal strength, $\mu_{\mathrm{Th}}\equiv \sigma_{\mathrm{Th}}/\sigma_\SM$, where $\sigma_{\mathrm{Th}}=\sigma_{\mathrm{Prod}}\times \mathrm{BR}$ includes the effects generated by the dimension-six operators. 
The theory predictions for the signal strengths are then used to build a test statistic in the form of  a $\log$-likelihood of a Gaussian distribution 
\begin{equation}
	\log(L) = -\frac{1}{2}\left[  (\vec{\mu}_{\mathrm{Exp}} -\vec{\mu} ) ^{T} \cdot \mathbf{V}^{-1} \cdot ( \vec{\mu}_{\mathrm{Exp}} -\vec{\mu} )\right]  .
	\label{eq:loglike}
\end{equation}
The covariance matrix~$\mathbf{V}$ is constructed from the experimental uncertainties $\delta \mu_{\mathrm{Exp}}$ and correlations\footnote{Correlations amongst channels of $< 10\%$ were ignored.},  as well as the theoretical uncertainties (scale, PDF, $\alpha_s$, \ldots).

The $\log$-likelihood of \eq{eq:loglike} was used together with flat priors~$ \pi(C_i)= const.$ in a Bayesian fit of the Wilson coefficients of interest.  A Markov chain Monte Carlo~(MCMC) using \texttt{pymc3}~\cite{Salvatier2016} was used to construct the posterior distribution. We use the \texttt{Arviz} Bayesian analysis package~\cite{arviz_2019} to extract the credible intervals (CIs) from the highest density posterior intervals~(HDPI) of the posterior distributions, where the intervals covering 95\% (68\%) of the posterior distribution are considered the 95\% (68\%) CIs. In the Gaussian limit, these  95\% (68\%) CIs should be interpreted as equivalent to the 95\%  (68\%) Frequentist  Confidence Level~(CL) two-sided bounds. To cross-check the MCMC Bayesian fit, a frequentist Pearson's $\chi^2$ fit was performed using~\texttt{iminuit}~\cite{James:1975dr,iminuit}, where the $\chi^2$ was taken to be 
\begin{equation}
	\chi^2 = - 2 \log(L).
\end{equation}
Both fit results agreed on the 95\%  and 68\% CI (or CL)  bounds.\footnote{In order to plot the multidimensional posterior distributions and the forest plots we have used a code based on ~\texttt{corner.py}~\cite{corner}, \texttt{pygtc}~\cite{Bocquet2016} and \texttt{zEpid}~\cite{paul_zivich_2019_3339870}.}
The code for the fit, experimental input and the analysis can be found in the repository \cite{GitHub}. The results of our calculations were also implemented in the {\tt HEPfit} code~\cite{DeBlas:2019ehy}, which was used for an independent cross-check of the results of the fits presented here. 
\par
In the theoretical predictions for the signal strengths, we will assume that the new physics corrections to the cross sections and the decay widths are linearised, i.e.
\begin{equation}
	\mu(C_\phi,C_i)=\frac{\sigma_{\rm Prod}(C_\phi,C_i) \times {\rm BR}(C_\phi,C_i)}{\sigma_{\rm Prod, SM}\times {\rm BR}_{\rm SM}} \approx 1+\delta \sigma(C_\phi,C_i)+\delta\Gamma(C_\phi,C_i)-\delta \Gamma_h(C_\phi,C_i),
	\label{linear-mu}
\end{equation}
with $\delta \sigma$, $\delta \Gamma$, $\delta \Gamma_h$ ($\Gamma_h$ denotes the Higgs total width) being the NLO corrections
, relative to the SM prediction as in \eq{eq:deltaR}, 
from the dimension-six operators with Wilson coefficients $C_\phi$ and $C_i$. Here, $C_i$ stands schematically for $C_{Qt}^{(1)}$, $C_{Qt}^{(8)}$, $C_{QtQb}^{(1)}$, $C_{QtQb}^{(8)}$, $C_{QQ}^{(1)}$, $C_{QQ}^{(3)}$ and $C_{tt}$.  As mentioned in the previous section, however, the sensitivity to $C_{QQ}^{(1),(3)}$ and $C_{tt}$ is rather small, typically below the theory uncertainty of the $t\bar{t}h$ calculation, and we will ignore these Wilson coefficients in the fits presented in this section. 

In particular, in \eq{linear-mu} all the corrections from the four-quark operators to the cross sections and decay widths are fully linearised in $1/\Lambda^2$. 
Given that current bounds on these operators are rather weak, one may wonder about the uncertainty in our fits associated to the truncation of the EFT.
Note that, since the four-quark operators only enter into the virtual corrections at NLO, Higgs production and decay contain only linear terms in $1/\Lambda^{2}$ in the corresponding Wilson coefficients, i.e.~the quadratic terms coming from squaring the amplitudes are technically of next-to-NLO. 
Hence, the leading quadratic effects in the signal strengths come from not linearising the corrections to the product $\sigma_{\rm Prod} \times {\rm BR}$~\!.  
We explicitly checked that, for the fits we presented in the next section, the difference between including the full expression of the signal strength or the linearised version in \eq{linear-mu} results in differences in the bounds at the $\lesssim10\%$ level. 
For the ${\cal O}_{\phi}$ operator, however, there is an additional contribution to the virtual corrections stemming from the wave function renormalisation of the Higgs field. The correction to a given production cross section or decay width, again denoted generically by $R$, is given by

\begin{equation}
\delta R_{\lambda_3}\equiv\frac{R_{\rm NLO}(\lambda_3)-R_{\rm NLO}(\lambda_3^{\rm{SM}})}{R_{\rm LO}}=-2\frac{C_{\phi}v^4}{\Lambda^2 m_h^2}C_1 + \left(-4\frac{C_{\phi}v^4}{\Lambda^2 m_h^2}+4\frac{C_{\phi}^2 v^8}{m_h^4\Lambda^4}\right) C_2 . \;\;\;\;\;
 \label{eq:degrassi}
\end{equation}
In \eq{eq:degrassi}, the coefficient $C_1$ corresponds to the contribution of the trilinear coupling to the single-Higgs processes at one loop, adopting the same notation as~\cite{Degrassi:2016wml}. The values of $C_1$ for the different processes of interest for this paper are given in~\autoref{App:numinput}. The coefficient $C_2$ describes universal corrections and is given by
\begin{equation}
	C_2=\frac{\delta Z_h}{1-\left(1-\frac{2 C_\phi v^4}{\Lambda^2 m_h^2}\right)^2 \delta Z_h}\,, \label{eq:C2}
\end{equation}
where the constant $\delta Z_h$ is the SM contribution from the Higgs loops to the wave function renormalisation of the Higgs boson,
\begin{equation}
	\delta Z_h =-\frac{9}{16}\,\frac{G_F m_h^2}{\sqrt{2}\pi^2}\left(\frac{2\pi}{3\sqrt{3}}-1\right).
\end{equation}
The coefficient $C_2$ thus introduces additional $\mathcal{O}(1/\Lambda^4)$ (and higher order) terms in $\delta R_{\lambda_3}$.  
In ref.~\cite{Degrassi:2016wml}, considering the $\kappa$ formalism, the full expression of \eq{eq:C2} is kept, while we define two different descriptions: one in which we expand $\delta R_{\lambda_3}$ up to linear order and an alternative scheme in which we keep also terms up to $\mathcal{O}(1/\Lambda^4)$ in the EFT expansion. We explicitly checked that keeping the full expression in \eq{eq:C2} and including terms up to $\mathcal{O}(1/\Lambda^4)$  in $C_2$ lead to nearly the same results in our fits.

\subsection{Fit to LHC Run-II data}

\par 
For the fit we have used inclusive Higgs data from the LHC Run II for centre-of-mass energy of $\sqrt{s} = 13$ TeV and  integrated luminosity of $ 139\, \mathrm{fb}^{-1}$ for ATLAS and  $ 137\,\mathrm{fb}^{-1}$ for CMS.  The experimental input is summarised in table~\ref{table:resHiggsExp} in~\autoref{App:numinput}.

\begin{figure}[htpb!]
	\vspace{-1 cm}
	\centering
	\includegraphics[width=0.75\linewidth]{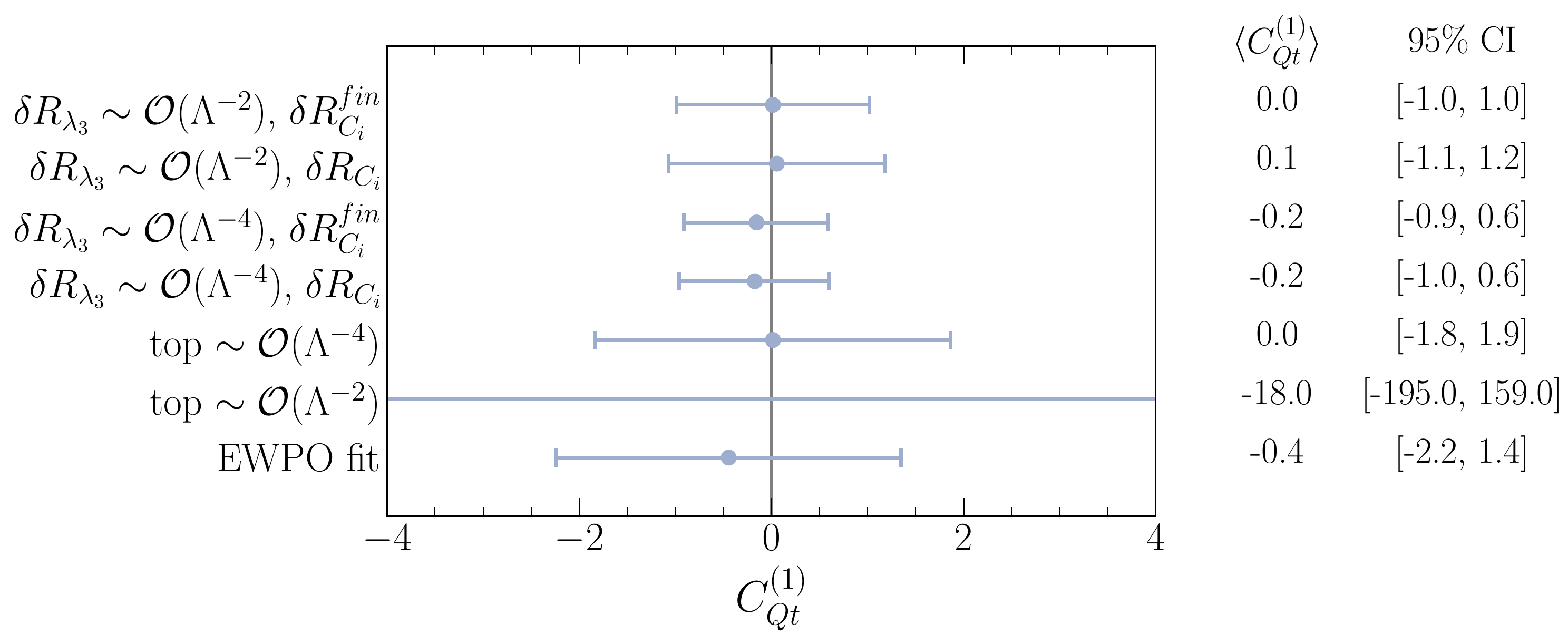}
	\includegraphics[width=.75\linewidth]{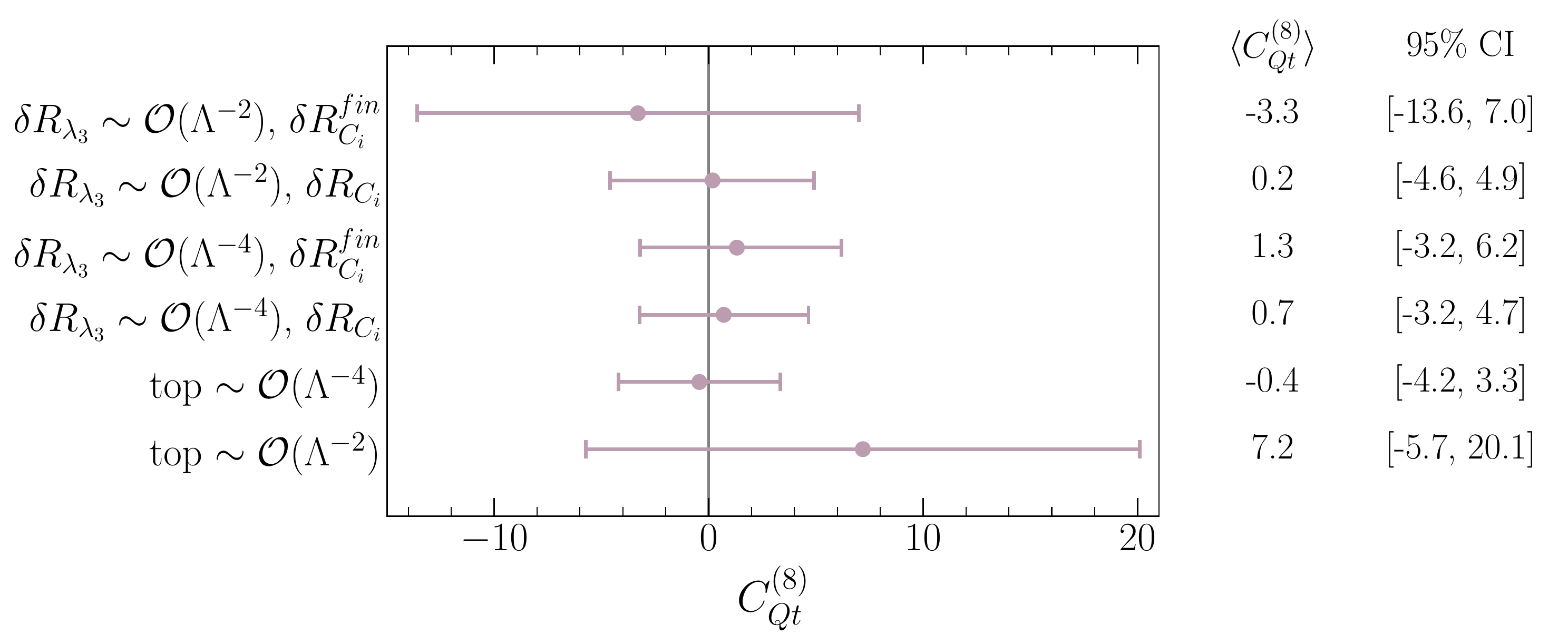} 
	\includegraphics[width=.75\linewidth]{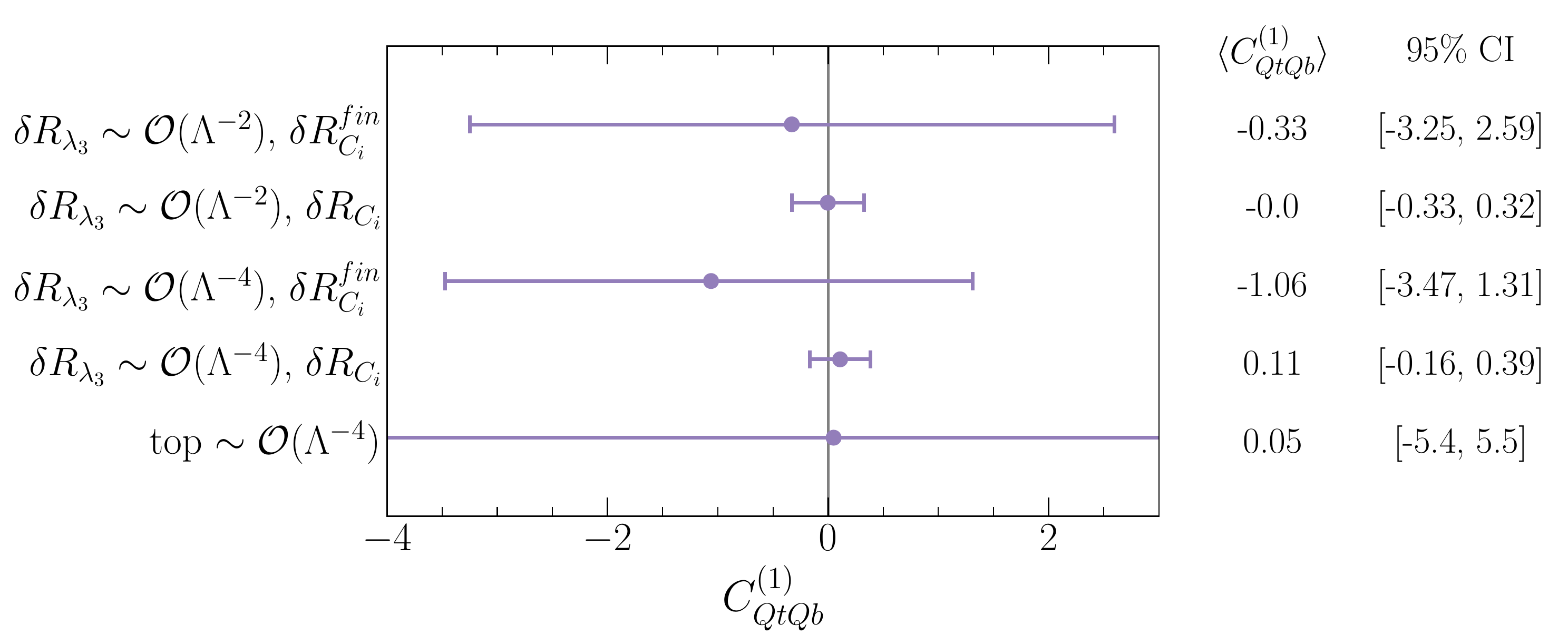}
	\includegraphics[width=.75\linewidth]{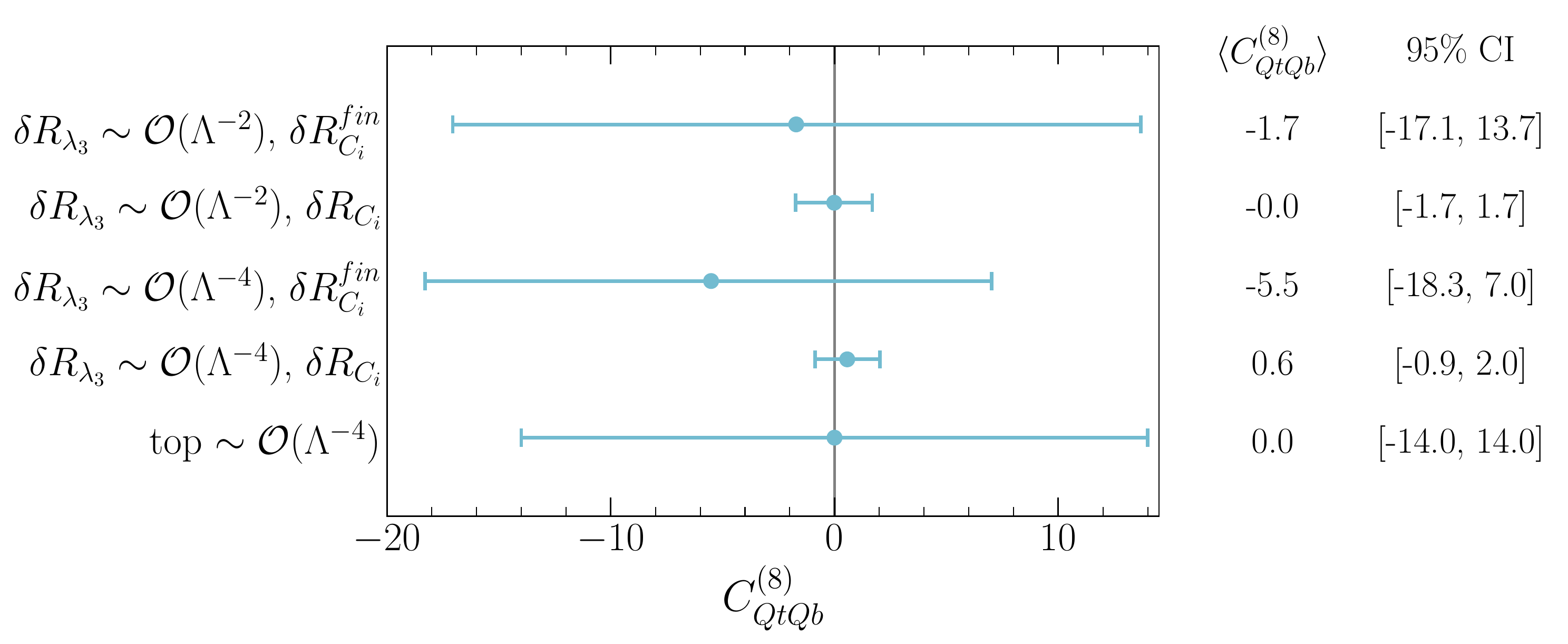}
	\caption{Forest plots illustrating the means and 95\% CIs of the posteriors built from the four-fermion Wilson coefficients with $C_\phi$ marginalised.  The plots confront also the truncation of the EFT at $\mathcal{O}(1/\Lambda^2)$ and $\mathcal{O}(1/\Lambda^4)$ of $\delta R_{\lambda_3}$, as defined in \eq{eq:degrassi}. The 95\% CI bounds stem from Higgs data. The last two rows for each operator show instead the limits obtained by a single parameter fit to top data, linear and quadratic. For the case of $C_{Qt}^{(1)}$, we also compare with the  results from the single-parameter fit using electroweak precision observables (EWPO) from \cite{Dawson:2022bxd}. The top data results are taken from~\cite{Ethier:2021bye} for $C_{Qt}^{(1),(8)}$ and~\cite{Hartland:2019bjb} for $C_{QtQb}^{(1),(8)}$. 
	\label{fig:summ4f} }
\end{figure}

In~\autoref{fig:summ4f} we show the limits of a two-parameter fit for various heavy four-quark Wilson coefficients $C_{i}$, marginalising over $C_\phi$. 
We confront them also with the limits obtained from fits to top data \cite{Ethier:2021bye, Ellis:2020unq, Hartland:2019bjb, Brivio:2019ius,DHondt:2018cww, Zhang:2017mls}. 
Note that, although our bounds do not come from a global fit, they can be compared with similar results from the fits to top data that assume that
only one operator is ``switched on'' at a time.
In these cases, we find that our new bounds are more stringent or at least comparable to the 95\% CI bounds on the  $C_{i}$ operators fit results from top data.  
For the case of ${\cal O}_{Qt}^{(1)}$, one can also see that the Higgs bounds are stronger than the results from a single-parameter fit to electroweak precision data, taken from the recent study in \cite{Dawson:2022bxd}.
We also note that, while the limits from top data show a large uncertainty from the EFT truncation\footnote{In particular, for the $C_{QtQb}^{(1),(8)}$ operator the references only calculate contributions of order $\mathcal{O}(1/\Lambda^4)$. (The fit considering only linear terms would result in bounds of order  $\mathcal{O}(10^4)$.) Hence, in this case, we only quote the quadratic bounds.}, even when only one operator is considered at a time, our NLO results for the four-quark operators are quite stable if one considers quadratic effects, as mentioned above. 
On the other hand, ~\autoref{fig:summ4f} also shows that the uncertainty associated to the EFT truncation of the effects of the $\mathcal{O}_{\phi}$ operator in the wave function renormalisation of the Higgs boson can be rather large.
Indeed, those effects can change the results by up to a factor of two, as it is the case for some of the $C_{Qt}^{(8)}$ limits. 
Furthermore, the plot displays the bounds for two different assumptions for the scale at which the operators are defined. The lines showing $\delta R_{C_i}^{fin}$ assume that there are only the corresponding four-quark operator and $\mathcal{O}_{\phi}$ at the electroweak scale\footnote{We neglect in this case the small running between the scales involved in the different processes included in the fit.}, while the line corresponding to $\delta R_{C_i}$ shows the limits assuming that the four-fermion operators (and $\mathcal{O}_{\phi}$) are the only ones at a scale $\Lambda=1\text{ TeV}$. 
We can infer from the fact that the bounds remain the same order of magnitude between $\delta R_{C_i}^{fin}$ and $\delta R_{C_i}$ that the inclusion of the finite terms for the operator $\mathcal{O}_{Qt}^{(1)}$, and to a less extent $\mathcal{O}_{Qt}^{(8)}$, is important if the new physics scale is not extremely high. Instead, for the operators $\mathcal{O}_{QtQb}^{(1),(8)}$ the bounds become much stronger when including the logarithmic piece, so we can conclude that in that case the finite piece is less relevant.

\begin{figure}[t!]
	\begin{center}
			\includegraphics[width=\linewidth]{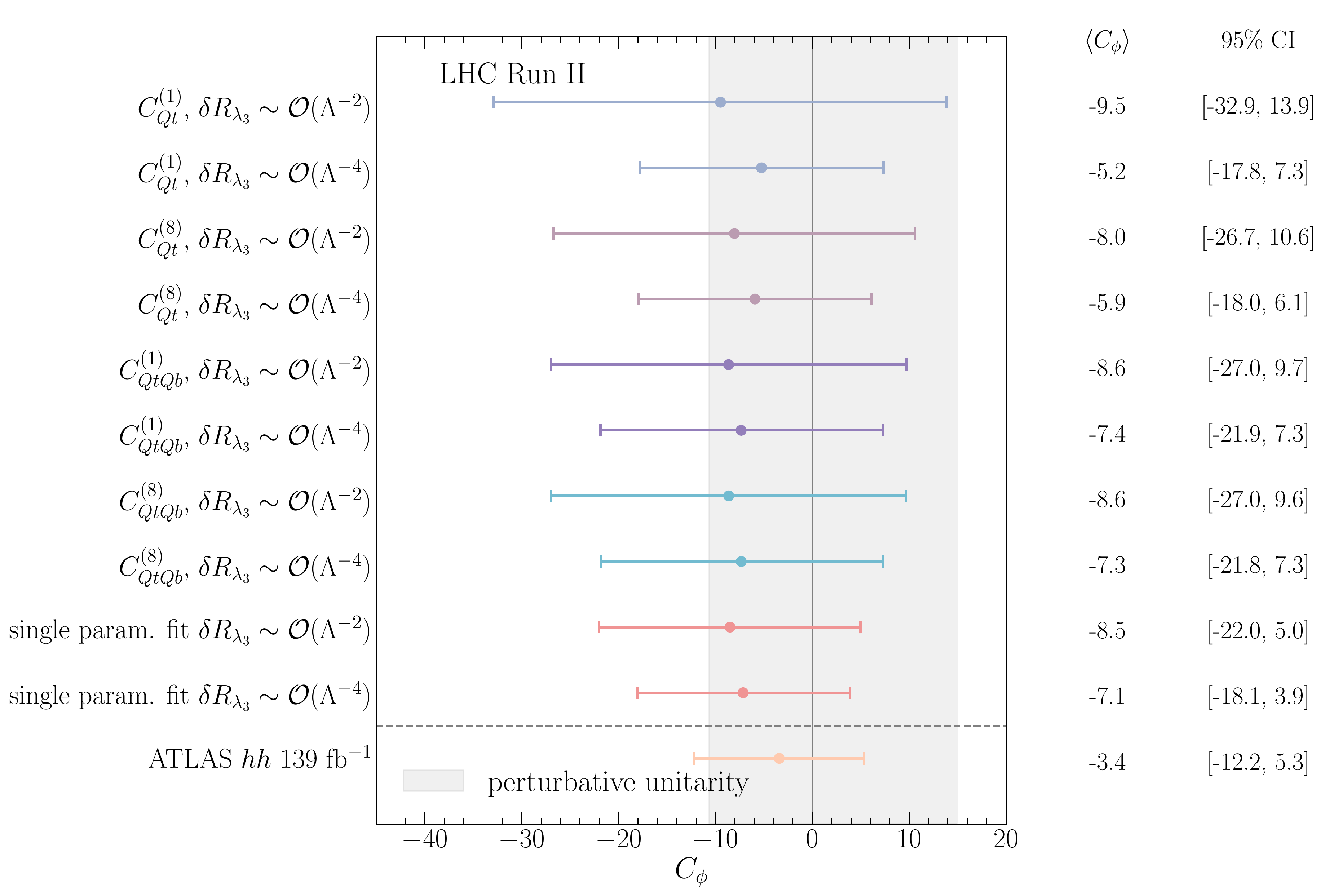}
	\end{center}
	\caption{A forest plot illustrating the means and 95\% CIs of the posteriors for $C_\phi$ from the two-parameter fits with the four-fermion operators marginalised. We compare the fit results for $C_\phi$ from full Run-II Higgs data keeping terms up to $\mathcal{O}(1/\Lambda^2)$ or $\mathcal{O}(1/\Lambda^4)$ in $\delta R_{\lambda_3}$.  For comparison, also the 95\% CI and means for the single parameter fit for $C_\phi$ with the same single-Higgs data are shown, as well as the bounds on $C_{\phi}$ from the $139$ fb$^{-1}$ search for Higgs pair production~\cite{ATLAS:2021jki}. The different four-fermion operators are assumed to be defined at $\Lambda$ and hence the fits include both the finite and logarithmic corrections in~\autoref{table:res4top}.	
	The horizontal grey band illustrates the perturbative unitarity bound~\cite{DiLuzio:2017tfn}. \label{fig:summcphi}  }
\end{figure}

In what follows, in all the fit results that we will present we will assume that the Wilson coefficients are always evaluated at the scale $\Lambda=1$ TeV. 
In \autoref{fig:summcphi} we show the limits on $C_\phi$ for various two-parameter fits, comparing the two different EFT truncations of $\delta R_{\lambda_3}$. 
The results thus correspond to the same two-parameter fits represented by the $\delta R_{C_i}$ lines in~\autoref{fig:summ4f}.
We also show the results from a single-parameter fit on $C_{\phi}$. 
For comparison, we show the ATLAS limits from full LHC Run-II Higgs pair production in the final state~$b\bar{b} \gamma \gamma$~\cite{ATLAS:2021jki} where we have translated the bounds from $\kappa_\lambda\equiv\lambda_3/\lambda_3^{\rm SM}$ to the SMEFT, keeping both linear and quadratic terms.  As can be seen, the limits on $C_{\phi}$ from single and double Higgs production are of more or less similar size when keeping terms up to $\mathcal{O}(1/\Lambda^4)$. In the single-Higgs fit, the overall limits become weaker if one keeps only terms up to $\mathcal{O}(1/\Lambda^2)$. Moreover, in all cases, the single-Higgs fit results remain questionable leading to limits that extend beyond the perturbative unitarity bound of ref.~\cite{DiLuzio:2017tfn} for $C_\phi<0$. For the Higgs pair production results, on the other hand, keeping linear or up to quadratic terms in the EFT expansion makes a negligible effect for the $C_\phi>0$ bound, while the bound weakens at linear order in $1/\Lambda^2$ for $C_\phi<0$~\cite{IML}, again potentially in conflict with perturbative unitarity.  

From~\autoref{fig:summcphi} we also see that the limits on $C_{\phi}$ become weaker in a two-parameter fit with the four-quark operators, indicating that in a proper global SMEFT fit also the loop effects of other weakly constrained operators, such as these, need to be accounted for. This will become even more apparent from the results of the four-parameter fit discussed in what follows.

One of the important aspects of multivariate studies is the correlation among variables. Apart from the two-parameter fits discussed above, here we also consider a four-parameter fit to $C_\phi$ plus the three directions in the four heavy-quark operator parameter space that the Higgs rates are
mostly sensitive to, i.e. neglecting $C_{QQ}^{(1),(3) }$ and $C_{tt}$, and trading $C_{QtQb}^{(1)}$ and $C_{QtQb}^{(8)}$ by $C_{QtQb}^{+}$.
\begin{figure}[t!]
	\begin{center}
		\vspace{-1.5cm}
		\includegraphics[width=.6\linewidth]{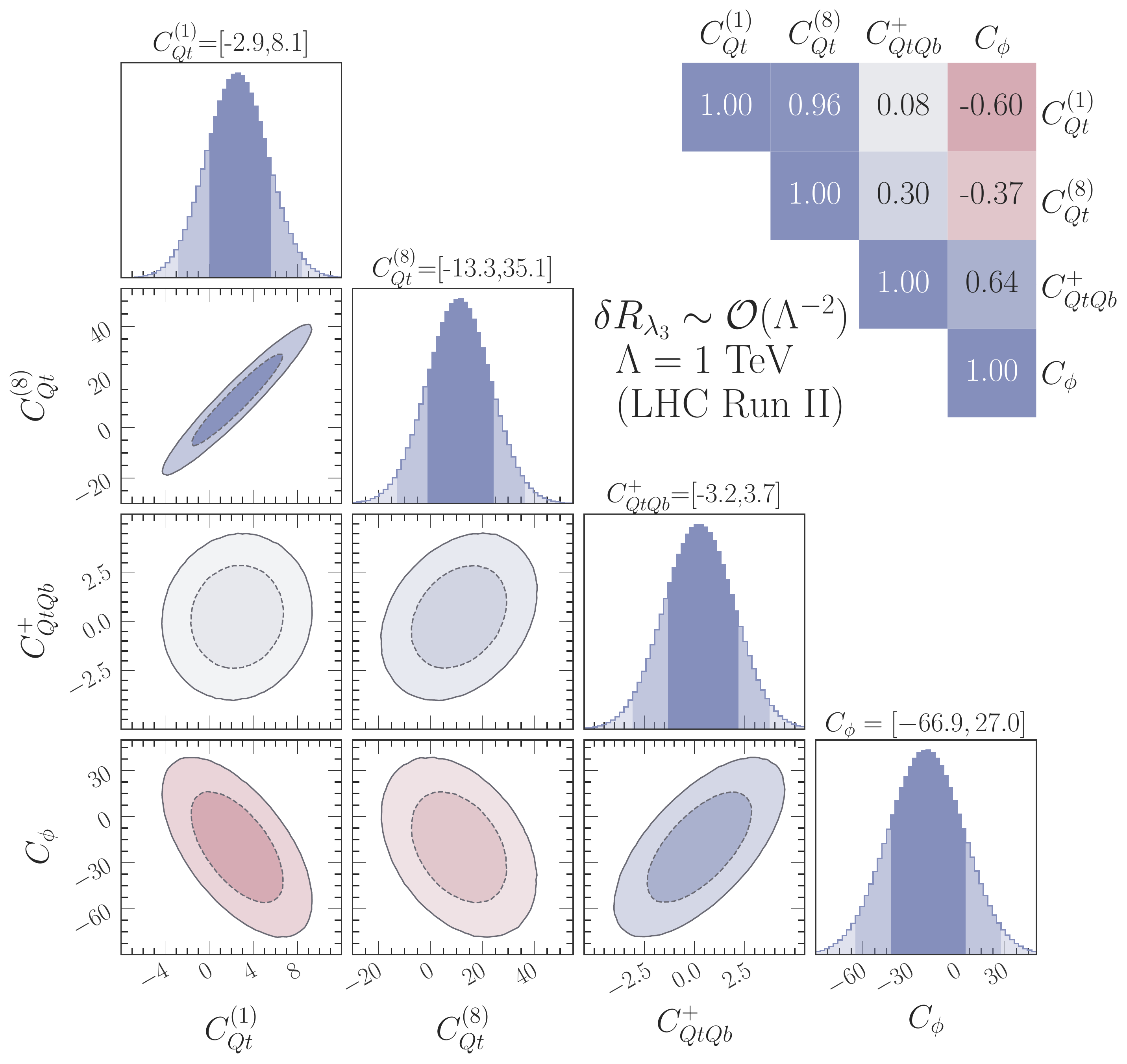}\\
		\includegraphics[width=.6\linewidth]{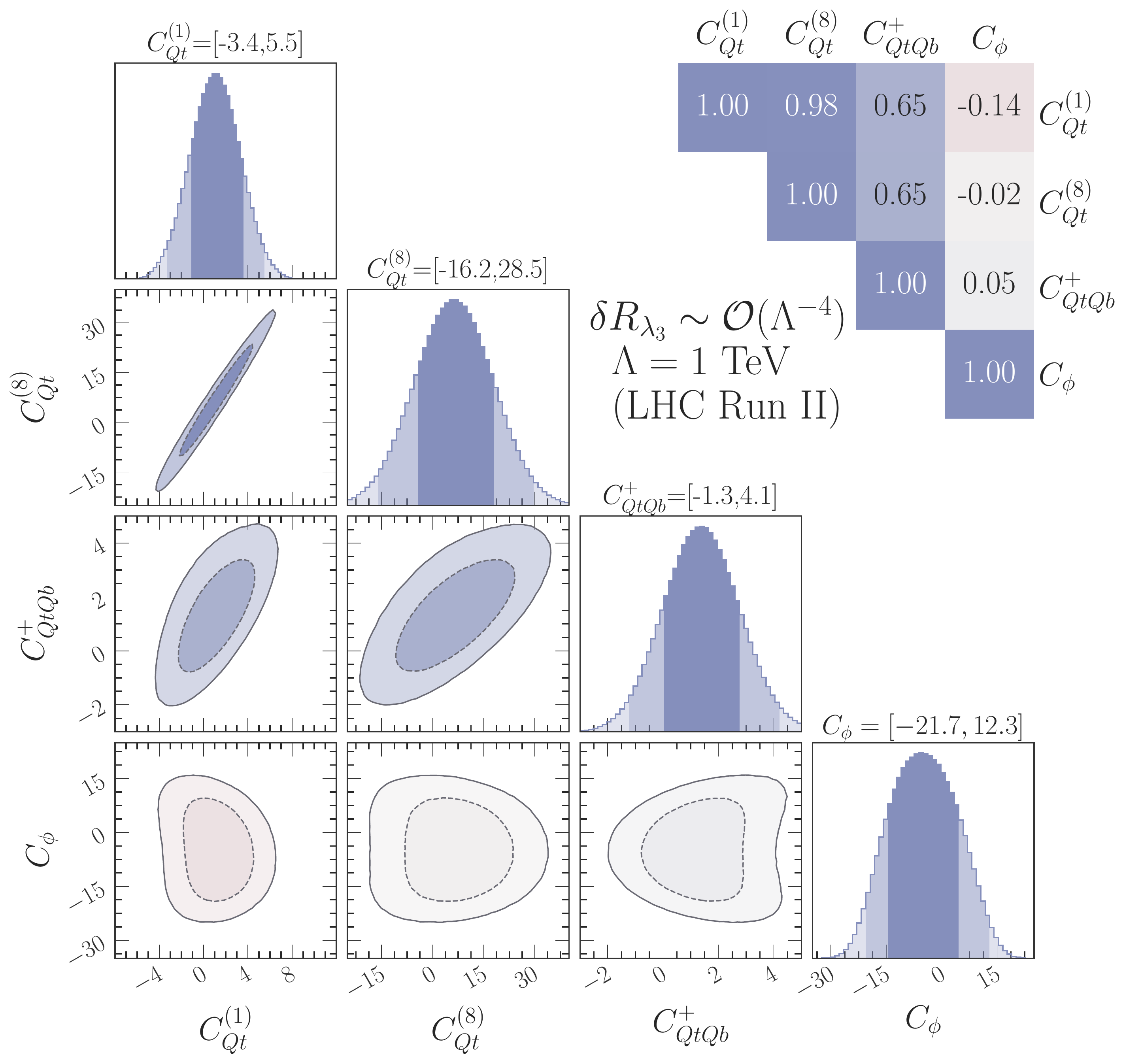}
		\vspace{-.5cm}
	\end{center}
	\caption{The marginalised 68\% and 95\% high density posterior contours and HDPI's for the four-parameter fits including the different four-quark Wilson coefficients and $C_\phi$. The numbers above the plots show the 95\% CI bounds while the correlations are given on the top-right corner. 
These limits correspond to values of the Wilson coefficients evaluated at the scale $\Lambda=1$ TeV. 
The upper panel shows the fit including terms up to $\mathcal{O}(1/\Lambda^2)$ in $\delta R_{\lambda_3}$, while the lower one shows the fit with including also  $\mathcal{O}(1/\Lambda^4)$.  \label{fig:4param} }
\end{figure}
When considering two- or four-parameter fits of $C_\phi$ and the four-heavy-quark Wilson coefficients, we observe a non-trivial correlation pattern amongst these coefficients. Figure~\ref{fig:4param} illustrates these correlation patterns clearly for the four-parameter fit. 
Focusing on the top panel, the results of the linear fit, 
we observe that the Wilson coefficients $C_{Qt}^{(1),(8) }$ are strongly correlated because, in analogy to $C_{QtQb}^{(1),(8) }$, they only appear in  certain linear combination whenever correcting the Yukawa coupling. However,  unlike $C_{QtQb}^{(1),(8) }$ they are not completely degenerate because the main part of the NLO correction to $t\bar t h$ does not contain the aforementioned linear combination. 
The four-parameter linear fit also reveals that the Wilson coefficients~$C_{Qt}^{(1),(8) }$ are somewhat decorrelated from ~$C_{QtQb}^{+}$.
Indeed, the fact that $t\bar{t}h$ and the Higgs decay $h \to b \bar b$ receive large NLO corrections only from  ~$C_{Qt}^{(1),(8) }$ and $C_{QtQb}^{(1),(8)}$, respectively, helps to separate both sets of operators. 
We also observe a relatively large correlation between the four-heavy-quark Wilson coefficients and $C_{\phi}$, though this depends on the $\delta R_{\lambda_3}$ truncation, and diminishes with the inclusion of the quadratic terms. 
As announced above, we observe again the impact of including the four-quark operators in the determination of the bound on $C_\phi$, which is much more pronounced in this four-parameter fit. In particular, the four-parameter linear fit yields a bound on $C_\phi$ $\sim 3$ times weaker that in the single $C_\phi$ fit.
In~\autoref{App:fitplots} we present similar correlation plots for various two-parameter fits, where the same behaviour of the change in the correlation with the inclusion of quadratic terms in ~$\delta R_{\lambda_3}$ is found.

\subsection{Prospects for HL-LHC}
\begin{figure}[t!]
	\begin{center}
		\includegraphics[width=0.75\linewidth]{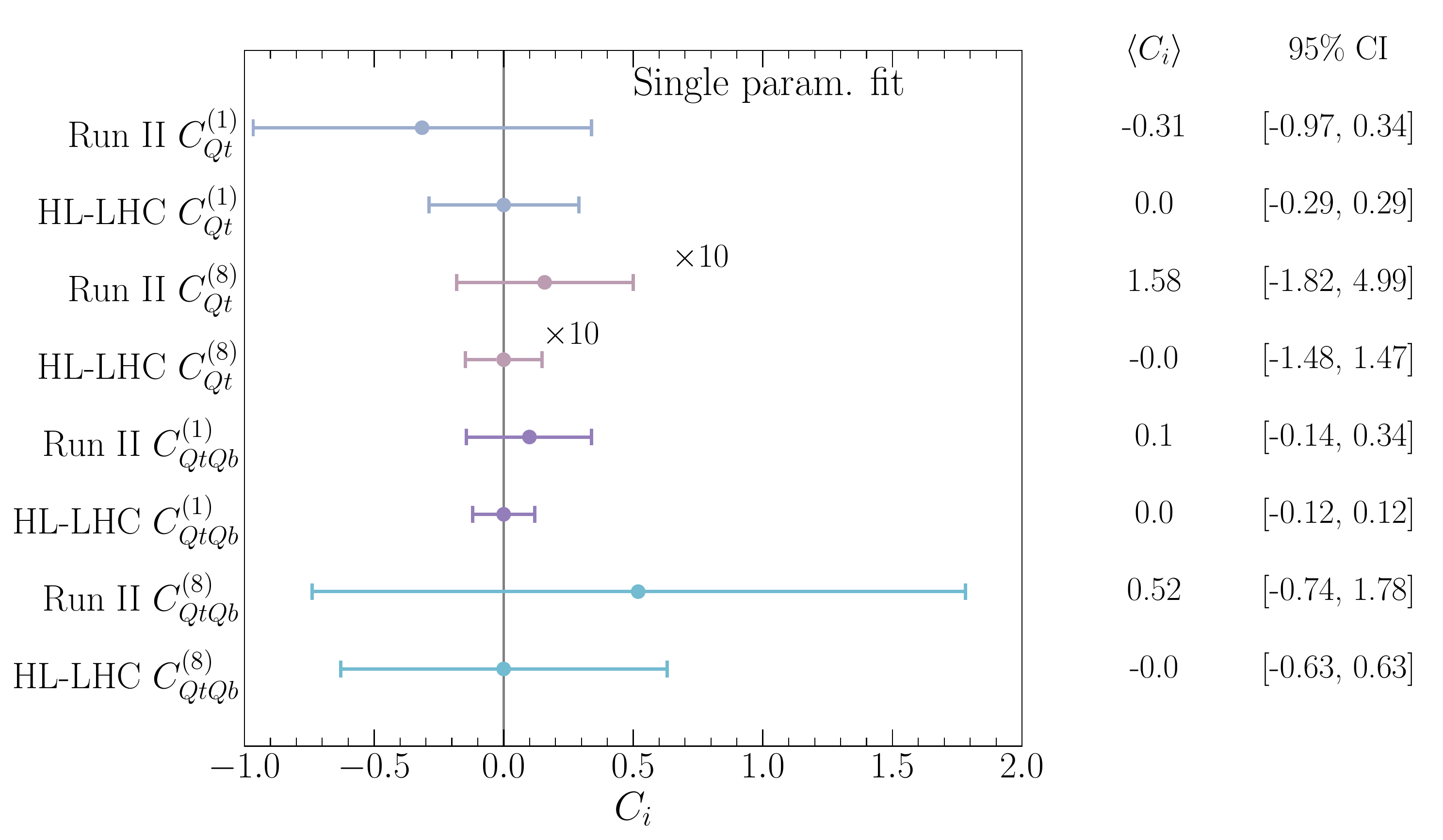}
	\end{center}
	\caption{ Results of a single parameter fit showing the improvement in constraining power of the HL-LHC over the current bounds from Run-II data. The limits correspond to values of the Wilson coefficients evaluated at the scale $\Lambda=1$ TeV. \label{fig:HLLHC} }
\end{figure}

\begin{figure}
	\begin{center}
		\includegraphics[width=0.75\linewidth]{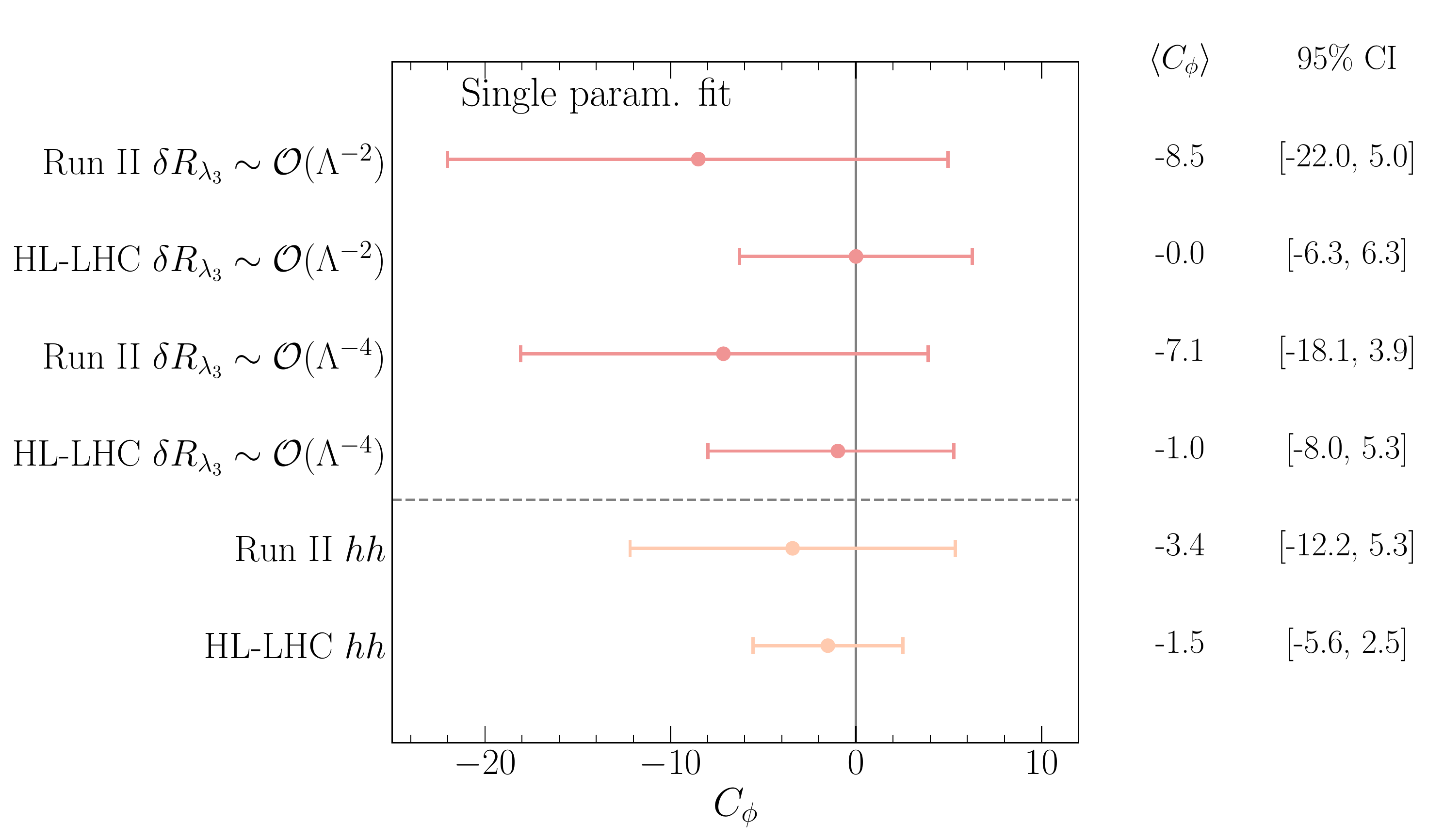}
	\end{center}
	\caption{A forest plot illustrating the means and 95\% CIs of the posterior for  $C_\phi$  from a single-parameter fit, showing also the differences in including terms of $\mathcal{O}(1/\Lambda^2)$ or up to $\mathcal{O}(1/\Lambda^4)$ in the definition of $\delta R_{\lambda_3}$. For comparison, also the limits and projections from searches for Higgs pair production are shown.  \label{fig:summcphihl-lhc}  }
\end{figure}
We now turn to examine the constraining power of the Higgs data that is expected to be collected at the HL-LHC. For this, we use the CMS projections for the single-Higgs signal strengths provided in refs.~\cite{CMS-PAS-FTR-18-011,twiki} for a centre-of-mass energy of $\sqrt{s}=14$ TeV and integrated luminosity of $ 3\, \mathrm{ab}^{-1}$. We use the projections for the S2 scenario explained in~\cite{Cepeda:2019klc}. These assume the improvement on the systematics that is expected to be attained by the end of the HL-LHC physics programme, and that theory uncertainties are improved by a factor of two with respect to current values. 
These projections are assumed to have their central values in the SM prediction with the total uncertainties summarised in~\autoref{table:resHiggsExp} in~\autoref{App:numinput}.\footnote{The correlation matrix for the S2 scenario can be found on the webpage~\cite{twiki}.} 

In~\autoref{fig:HLLHC} we confront the results of single-parameter fits to Run-II data for each of the four-quark operators with the projections for the HL-LHC. For all the four-quark operators the constraining power of the HL-LHC is roughly a factor two better than the current bounds we could set from single-Higgs data, with a slightly lower improvement for the operators $\mathcal{O}_{QtQb}^{(1),(8)}$ compared to $\mathcal{O}_{Qt}^{(1),(8)}$.
In~\autoref{fig:summcphihl-lhc} we show the limits on $C_{\phi}$ in a single parameter fit for Run-II and the projections for the HL-LHC
including corrections in $\delta R_{\lambda_3}$ up to order $\mathcal{O}(1/\Lambda^2)$ or $\mathcal{O}(1/\Lambda^4)$. 
As expected, the inclusion of terms of $\mathcal{O}(1/\Lambda^4)$ makes a less pronounced difference for the HL-LHC projection compared to the Run-II results. 
Our results are very similar to the projections presented in a $\kappa_{\lambda}$ fit in \cite{DiMicco:2019ngk}. 
We confront this also with data from searches for Higgs pair production $139$ fb$^{-1}$ \cite{ATLAS:2021jki}  and HL-LHC projections~\cite{CMS:2018ccd} on Higgs pair production, showing that Higgs pair production will still allow to set stronger limits on $C_{\phi}$.

\section{Summary and discussion \label{sec:conclusion}}

In this paper, we have computed the NLO corrections to Higgs observables induced by third generation four-quark operators relevant for single-Higgs production and decay at the LHC. 
Our results show that such processes are sensitive to the all possible chiral structures for the third generation four-quark operators in the dimension-six SMEFT, but in different degrees. 
Operators with different chiralities are, for instance, the only ones that can contribute to Higgs production via gluon fusion, and the decay of the Higgs boson to gluons, photons and bottom quarks pairs. The latter are particularly sensitive to the top-bottom operators $\mathcal{O}_{QtQb}^{(1),(8)}$, which then also significantly affect the total decay width. In the associate production of a Higgs boson with a top quark pair, on the other hand, a priori all the third generation four-quark operators enter.
Sensitivity to four-quark operators where all fields have the same chirality, however, is only possible for very large values of the corresponding effective interactions, in a way that they can generate contributions beyond the size of current theory uncertainties, but possibly in a regime in conflict with the EFT expansion. Contributions from ``mixed'' top-bottom operators are also highly suppressed.
The $t\bar{t}h$ process is, in fact, particularly important in setting limits on the four-quark operators $\mathcal{O}_{Qt}^{(1)}$ and $\mathcal{O}_{Qt}^{(8)}$, due to the comparatively large NLO corrections they induce in this process with respect to others. It also breaks a degeneracy among the Wilson coefficients of those two operators, which always appear in a single combination for all other processes. 
\par
To illustrate the constraining power of single-Higgs processes in bounding these four-quark operators, we performed several simplified fits of these interactions to Higgs data and find that the resulting limits from our fits are, in some cases, comparable or better than similar results obtained from top data \cite{Ethier:2021bye, Hartland:2019bjb}.
\par
We have also performed fits including the above-mentioned four-quark operators and the operator $\left(\phi^\dagger \phi\right)^3$, that modifies the Higgs potential and the trilinear Higgs self-coupling. Due to the lack of powerful constraints from top data, the inclusion of the four-fermion operators diminishes the power of setting limits on the trilinear Higgs self-coupling
from single-Higgs observables. 
From our analysis we conclude that, in the absence of strong direct bounds on the third-generation four-quark operators, these should be included into a global fit to Higgs data, when attempting to obtain model-independent bounds on the trilinear Higgs self-coupling. The results of our calculations are presented such that they can be easily used by the reader in truly global fits including all other interactions entering at the LO. 
We leave this, as well as the inclusion of differential Higgs data, to future work.

Finally, we also illustrated the increase in constraining power expected during the high-luminosity phase of the LHC by presenting the HL-LHC projections for single-parameter fits.  

Moving beyond hadron colliders, it must be noted that the interplay between the Higgs trilinear and four heavy-quark operators in Higgs processes is expected to be less of an issue at future leptonic Higgs factories, such as the FCC-ee~\cite{FCC:2018byv,FCC:2018evy}, ILC~\cite{Bambade:2019fyw,LCCPhysicsWorkingGroup:2019fvj}, CEPC~\cite{An:2018dwb,CEPCStudyGroup:2018ghi} or CLIC~\cite{CLICdp:2018cto,deBlas:2018mhx}. At these machines, the effects of $C_\phi$ are still ``entangled'' with those
of the four-fermion operators in the Higgs rates, but only through the decay process, i.e.~via the contributions to the BRs. However, Higgs production is purely electroweak, namely via Higgs-strahlung ($Zh$: $e^+ e^- \to Zh$) or $W$ boson fusion, and receives no contributions from the four-quark operators at the same order in perturbation theory where $C_\phi$ modifies these processes, i.e.~NLO. 
Moreover, at any of these future $e^+ e^-$ Higgs factories there is the possibility of obtaining a sub-percent determination of the total $Zh$ cross section at $e^+ e^-$ colliders, by looking at events recoiling against the $Z$ decay products with a recoil mass around $m_h$. This observable is therefore completely insensitive to the four-quark operators, while still receiving NLO corrections from $C_\phi$. 
Although, in practice, in a global fit one needs to use data from all the various Higgs rates at two different energies to constrain all possible couplings entering at LO in the Higgs processes and also obtain a precise determination of $C_\phi$ ~\cite{DiVita:2017vrr}, the previous reasons should facilitate the interpretation of the single-Higgs bounds on the Higgs self-coupling at $e^+e^-$ machines.

We conclude this paper with a few words on the relevance of the results presented here when interpreted from the point of view of specific models of new physics. 
In particular, one important question is {\it are there models where one expects large contributions to four-top operators while all other interactions entering in Higgs processes are kept small?}
Indeed, large contributions to four-top operators can be expected in various BSM scenarios.\footnote{Generically, models where four-top interactions are much larger than four-fermion operators of the first and second generation can be easily conceived from some UV dynamics coupling mostly to the third generation of quarks hence respecting the Yukawa hierarchies.} For instance, in Composite Higgs Models, in which the top quark couples to the strong dynamics by partial compositeness, one expects on dimensional grounds that some of the four-top quark operators are of order $1/f^2$, where $f$ indicates the scale of strong dynamics \cite{Banelli:2020iau}. 
By its own nature, however, Composite Higgs models also predict sizeable contributions to the single-Higgs couplings $\sim 1/f^2$. While, in general, sizeable modifications of the Higgs interactions are typically expected in models motivated by ``naturalness'', this is not necessarily the case in other scenarios. 
It is indeed possible to think of simple models where modifications of the Higgs self-interactions or contributions to four-quark operators are the only corrections induced by the dimension-six interactions at tree level, see~\cite{deBlas:2017xtg}.
Thinking, for instance, in terms of scalar extensions of the SM, there are several types of colored scalars whose tree-level effects at low energies can be represented by four-quark operators only, e.g.~for complex scalars in the $(6,1)_{\frac 1 3}$ and $(8,2)_{\frac 1 2}$ SM representations ($\Omega_1$ and $\Phi$ in the notation of \cite{deBlas:2017xtg}).  If these colored states are the only moderately heavy new particles, our results can provide another handle to constrain such extensions. 
One must be careful, though, as a consistent interpretation of our results for any such models would require to include higher-order corrections in the matching to the SMEFT. At that level, as shown e.g.~by the recent results in \cite{Anisha:2021hgc}, multiple contributions that modify Higgs processes at LO are generated at the one-loop level, and are therefore equally important as the NLO effects of the (tree-level) generated four-quark operators.\footnote{Furthermore, given that some SMEFT interactions induce tree-level contributions to Higgs processes that in the SM are generated at the loop level, e.g. ${\cal O}_{\phi G}$ in gluon fusion, a consistent interpretation in terms of new physics models may require to include up to two-loop effects in the matching for such operators, for which there are currently no results nor tools available.}
In any case, one must note that, even if similar size contributions to single-Higgs processes are generated, the four-top or Higgs trilinear effects can provide complementary information on the model.
For instance, in some of the most common scalar extensions of the SM, with an extra Higgs doublet, $\varphi\sim (1,2)_{\frac 12}$, tree-level contributions to some of the four-heavy-quark operators discussed in this paper are generated together with modifications on the Higgs trilinear self-coupling.
These two effects are independent but they are both correlated with the, also tree level, modifications of the single-Higgs couplings. Essentially, the LO effects on Higgs observables are proportional to $\lambda_\varphi y_\varphi^f$, where $\lambda_\varphi $ is the scalar interaction strength of the $(\varphi^\dagger \phi)(\phi^\dagger \phi)$ operator and $y_\varphi^f$ the new scalar Yukawa interaction strength, whereas the NLO effects are proportional to the square of each separate coupling. Hence, these effects might help to resolve (even if only weakly) the flat directions in the model parameter space that would appear in a LO global fit.
At the end of the day, for a proper interpretation of the SMEFT results in terms of the widest possible class of BSM models, all the above simply remind us of the importance of being global in SMEFT analyses, to which our work contributes by including effects in Higgs physics that enter at the same order in perturbation theory as modifications of the Higgs self-coupling.

\subsection*{Acknowledgements}
We would like to thank Ayan Paul for providing functions used in~\cite{Grojean:2020ech} that facilitated making some of the plots shown in this paper. L.A. thanks the computing resources provided by DESY and the INFN, Sezione di Padova, for hospitality during the final stage of this work.  R.G. would like to thank Pier Paolo Giardino, Ken Mimasu, Paride Paradisi and Eleni Vryonidou for interesting discussions on $C_{\phi}$ and the four-fermion operators considered.  L.A. 's research is supported by the Deutsche Forschungsgemeinschaft (DFG) - Projektnummer 417533893/GRK2575  ``Rethinking Quantum Field Theory". The work of J.B. has been supported by the FEDER/Junta de Andaluc\'ia project grant P18-FRJ-3735.

\clearpage
\newpage

\appendix

\section{Numerical input \label{App:numinput} }

  Aside from our own calculations of the four-quark operator effects in single-Higgs rates, we have also used in our fits the dependence on the Higgs trilinear self-coupling of the NLO corrections to the same processes, which was calculated in ref.~\cite{Degrassi:2016wml}. Here we give them in~\autoref{table:resch}, translating the $\kappa_\lambda$ dependence 
in terms of~$C_\phi$,
\begin{equation}
\delta \kappa_\lambda = -2 \frac{C_\phi v^4}{m_h^2\Lambda^2},
\end{equation}
and assuming $\Lambda= 1$ TeV.

\begin{table}[h]
\centering
\begin{tabular}{ccc}
\toprule
Process&$C_1$ & $\delta R_{C_\phi}^{fin}$ \\
\midrule
ggF/ $gg\to h$  & $6.60\cdot 10^{-3}$ &  $-3.10\cdot 10^{-3}$ \\
$t\bar{t}h$   \textcolor{Mahogany}{13 TeV} & $3.51\cdot 10^{-2}$  &  $-1.64\cdot 10^{-2}$  \\
$t\bar{t}h$   \textcolor{Mahogany}{14 TeV} & $3.47\cdot 10^{-2}$  &  $-1.62\cdot 10^{-2}$ \\
$h\to \gamma \gamma$   & $4.90\cdot 10^{-3}$ &  $-2.30\cdot 10^{-3}$ \\
$h\to b\bar{b}$ & 0.00&0.00  \\
$h\to W^+ W^-$  & $7.30\cdot 10^{-3}$  & $-3.40\cdot 10^{-3}$ \\
$h\to Z Z$      & $8.30\cdot 10^{-3}$&  $-3.90\cdot 10^{-3}$ \\
$pp\to Zh$   \textcolor{Mahogany}{13 TeV}    & $1.19\cdot 10^{-2}$& $-5.60\cdot 10^{-3}$ \\
$pp\to Zh$    \textcolor{Mahogany}{14 TeV}    &$1.18\cdot 10^{-2}$&  $-5.50\cdot 10^{-3}$ \\
$pp\to W^\pm h$  & $1.03\cdot 10^{-2}$ &    $-4.80\cdot 10^{-3}$\\
VBF              & $6.50\cdot 10^{-3}$&   $-3.00\cdot 10^{-3}$ \\
$ h \to 4 \ell$       &     $8.20\cdot 10^{-3}$  &   $-3.80\cdot 10^{-3}$ \\
\bottomrule
\end{tabular}
\caption{The relative correction dependence on $C_\phi$ for single-Higgs processes, taken from~\cite{ Degrassi:2021uik}. The $C_1$ coefficients are to be used in eq.~\eqref{eq:degrassi}, while for a direct  comparison with the effect of the four-fermion operators, we quote the translated effect~$\delta R_{C_\phi}^{fin}$, which can be used directly in eq.~\eqref{eq:deltar}. If the value of $\sqrt{s}$ is not indicated the effect is the same for both $13$ and $14$ TeV.
}
\label{table:resch}
\end{table}

We also provide in this appendix the experimental measurements of the signal strengths at the LHC Run II and the CMS projections for the HL-LHC (scenario S2, see \cite{Cepeda:2019klc}) that we used in the fits in this paper. These inputs are summarised in~\autoref{table:resHiggsExp}.
 
 \begin{table}[ht!]
\centering
\vspace{-1 cm}
{ \footnotesize 
	{\renewcommand{\arraystretch}{0.75 }%
\begin{tabular}{clccc}
\toprule
\toprule
\multirow{5}{*}{ {\normalsize Production}}  &\multirow{5}{*}{ {\normalsize Decay}}&\multicolumn{2}{c}{ $\mu_{\mathrm{Exp}} \pm \delta \mu_{\mathrm{Exp}}$  (symmetrised)} &\multirow{5}{*}{ {\normalsize Ref.}} \\
&   & { \bf     \scriptsize           LHC Run II}&{ \bf  \scriptsize HL-LHC}&   \\
\cmidrule(r){3-4}
&   & { \scriptsize                   CMS $137 \, \mathrm{fb}^{-1} $}&  \multirow{2}{*}{CMS $3 \, \mathrm{ab}^{-1}$}&   \\
&   &  { \CG \scriptsize                   ATLAS $139 \, \mathrm{fb}^{-1} $} & &  \\
\midrule
\midrule
\multirow{ 13}{*}{ \normalsize ggF}         & \multirow{2}{*}{$h\to \gamma  \gamma$} & { \scriptsize                  $0.99 \pm 0.12$}& \multirow{2}{*}{$1.000\pm 0.042$}& \multirow{2}{*}{\cite{ATLAS:2020qdt,CMS:2021kom,CMS-PAS-FTR-18-011}}\\
                                           &                                                          &{ \scriptsize                   \CG $1.030 \pm 0.110$}&& \\ 
                                           \cmidrule(r){2-5}
                                    &  \mr{$h\to Z Z^*$}          & { \scriptsize                  $0.985 \pm 0.115$}&\multirow{2}{*}{$1.000 \pm 0.040$}&\multirow{7}{*}{\cite{ATLAS:2020qdt,CMS:2020gsy,CMS-PAS-FTR-18-011}}  \\
                                     &                                                      &{ \scriptsize                   \CG $0.945 \pm 0.105$}&& \\
                                     \cmidrule(r){2-4}
                                    &\mr{$h\to W W^*$}         & { \scriptsize                  $1.285 \pm 0.195$} &\mr{ $1.000 \pm 0.037$} &\\
                                    & &                                            { \scriptsize                   \CG$1.085 \pm 0.185$} & &\\
                                                                         \cmidrule(r){2-4}
                                    &\mr{$h\to \tau^+\tau^- $}         & { \scriptsize                  $0.385 \pm 0.385$} &\mr{ $1.000 \pm 0.055$} &\\
                                 & &                                            { \scriptsize                   \CG$1.045 \pm 0.575$} & &\\
                                 \cmidrule(r){2-5}

                                  &\mr{$h\to  b \bar b$}      & { \scriptsize                 $2.54 \pm 2.44$} &\mr{ $1.000 \pm 0.247$} &\mr{\cite{CMS:2020gsy,CMS-PAS-FTR-18-011}}\\
                               & &                                            { \scriptsize                   \CG--} & &\\
                                 \cmidrule(r){2-5}
                                  &\mr{$h\to  \mu^+ \mu^-$}      & { \scriptsize      $0.315 \pm 1.815$} &\mr{ $1.000 \pm 0.138$} &\mr{\cite{CMS:2020gsy,CMS-PAS-FTR-18-011} }\\
& &                                            { \scriptsize                   \CG--} & &\\
\midrule
\midrule
\multirow{13}{*}{ \normalsize VBF}      
										&\mr{$h\to \gamma  \gamma$}         & { \scriptsize                  $1.175 \pm 0.335$ } &\mr{ $1.000 \pm 0.128$} & \mr{\cite{ATLAS:2020qdt,CMS:2021kom,CMS-PAS-FTR-18-011}}\\
										& &                                           { \scriptsize                   \CG$1.325 \pm 0.245$} & &\\
\cmidrule(r){2-5}
                                     &\mr{$h\to Z Z^*$ }         & { \scriptsize                  $0.62 \pm 0.41$ } &\mr{ $1.000 \pm 0.134$} & \multirow{7}{*}{\cite{ATLAS:2020qdt,CMS:2020gsy,CMS-PAS-FTR-18-011}}\\
                                    & &                                            { \scriptsize                   \CG$1.295 \pm 0.455$} & &\\
                                                                                             \cmidrule(r){2-4}

                                   &\mr{$h\to W W^*$}         & { \scriptsize                  $0.65 \pm 0.63$ } &\mr{ $1.000 \pm 0.073$} & \\
                                    & &                                            { \scriptsize                   \CG$0.61 \pm 0.35$} & &\\
 \cmidrule(r){2-4}
                                   &\mr{$h\to \tau^+\tau^-$}         & { \scriptsize                  $1.055 \pm 0.295$ } &\mr{ $1.000 \pm 0.044$} & \\
& &                                            { \scriptsize                   \CG$1.17 \pm 0.55$} & &\\
\cmidrule(r){2-5}
                                    &\mr{$h\to  b \bar b$}         & { \scriptsize                   -- } &\mr{--} & \mr{\cite{ATLAS:2020qdt}}\\
                                    & &                                            { \scriptsize                   \CG$3.055 \pm 1.645$} & &\\
                                    
\cmidrule(r){2-5}
                                    &\mr{$h\to  \mu^+ \mu^-$}      & { \scriptsize               $3.325 \pm 8.075$} &\mr{ $1.000 \pm 0.540$} &\mr{\cite{CMS-PAS-FTR-18-011}}\\
 & &                                            { \scriptsize                   \CG--} & & \\                                   
\midrule
\midrule
\multirow{10}{*}{ \normalsize  $t\bar t h$} 
&\mr{$h\to \gamma  \gamma$}         & { \scriptsize                $1.43 \pm 0.30$ } &\mr{$1.000 \pm 0.094$} & \mr{ \cite{ATLAS:2020qdt,CMS:2021kom,CMS-PAS-FTR-18-011} }\\
& &                                            { \scriptsize                   \CG$0.915 \pm 0.255$} & &\\

\cmidrule(r){2-5}

&\multirow{3}{*} {$h\to V V^*$}         & { \scriptsize              $0.64 \pm 0.64$({\color{Mahogany}$ZZ^*$}) } &{ \scriptsize   $1.000 \pm 0.246$ ({\color{Mahogany}$ZZ^*$}) } & \multirow{8}{*}{\cite{ATLAS:2020qdt,CMS:2020gsy,CMS-PAS-FTR-18-011}}  \\
& &                                            { \scriptsize                   $0.945\pm 0.465$ ({\color{Mahogany} $W W^*$})} & { \scriptsize   $1.000 \pm 0.097$ ({\color{Mahogany} $W W^*$})} &\\
& &                                            { \scriptsize                   \CG $1.735 \pm 0.545$} & { \scriptsize   --}&\\
\cmidrule(r){2-4}                                    

&\mr{$h\to \tau^+\tau^-$}         & { \scriptsize                $0.845 \pm 0.705$} &\mr{ $1.000 \pm 0.149$} & \\
& &                                            { \scriptsize                   \CG $1.27 \pm 1.0$} & &\\
\cmidrule(r){2-4}                                    

&\mr{$h\to  b \bar b$}         & { \scriptsize                 $1.145 \pm 0.315$} &\mr{ $1.000 \pm 0.116$} & \\
& &                                            { \scriptsize                   \CG $0.795 \pm 0.595$} & &\\                                                        
\midrule
\midrule
\multirow{9}{*}{ \normalsize $Vh$}

&\mr{$h\to \gamma  \gamma$}         & { \scriptsize   $0.725 \pm 0.295$ } &{ \scriptsize   $1.000 \pm 0.233$ ({\color{Mahogany}$Zh$}) } & \multirow{2}{*}{ \cite{ATLAS:2020qdt,CMS:2021kom,CMS-PAS-FTR-18-011}  }  \\
& &                                            { \scriptsize                   \CG $1.335 \pm 0.315$} & { \scriptsize   $1.000 \pm 0.139$ ({\color{Mahogany} $W^\pm h$})} &\\
\cmidrule(r){2-5}           
                                    
&\mr{$h\to Z Z^*$}         & { \scriptsize   $1.21 \pm 0.85$ } &{ \scriptsize   $1.000 \pm 0.786$ ({\color{Mahogany}$Zh$}) } & \multirow{2}{*}{ \cite{ATLAS:2020qdt,CMS:2020gsy,CMS-PAS-FTR-18-011}  }  \\
& &                                            { \scriptsize                   \CG $1.635 \pm 1.025$} & { \scriptsize   $1.000 \pm 0.478$ ({\color{Mahogany} $W^\pm h$})} &\\
\cmidrule(r){2-5}           
                                  
 &\mr{$h\to W W^*$}         & { \scriptsize   $1.850\pm 0.438$ } &{ \scriptsize   $1.000 \pm 0.184$ ({\color{Mahogany}$Zh$}) } & \multirow{2}{*}{  \cite{CMS:2021ixs,CMS-PAS-FTR-18-011} }  \\
 & &                                            { \scriptsize                   \CG --} & { \scriptsize   $1.000 \pm 0.138$ ({\color{Mahogany} $W^\pm h$})} &\\
 \cmidrule(r){2-5}           
 &\mr{$h\to  b \bar b$}         & { \scriptsize  -- } &{ \scriptsize   $1.000 \pm 0.065$ ({\color{Mahogany}$Zh$}) } & \multirow{2}{*}{  \cite{ATLAS:2020qdt,CMS-PAS-FTR-18-011} }  \\
& &                                            { \scriptsize                   \CG $1.025 \pm 0.175$} & { \scriptsize   $1.000 \pm 0.094$ ({\color{Mahogany} $W^\pm h$})} &\\

                                    
\midrule
\midrule
\multirow{2}{*}{ \normalsize $Zh$ { \scriptsize {\color{Mahogany} CMS     }   }}    & $h\to \tau^+\tau^- $ & $1.645 \pm 1.485$&\multirow{5}{*}{--} &\multirow{5}{*}{ \cite{CMS:2020gsy} }  \\
& $h\to  b \bar b$       &$0.94 \pm 0.32$&&\\                         
 \cmidrule(r){2-3}    
\multirow{2}{*}{ \normalsize $W^\pm h${ \scriptsize {\color{Mahogany} CMS     }   }}           & $h\to \tau^+\tau^- $ &$3.08 \pm 1.58$&&\\
& $h\to  b \bar b$      & $1.28 \pm 0.41$&&\\                  
\midrule
\midrule
\end{tabular}
}
}
\caption{The experimental measurements of single-Higgs observables from the LHC Run II and projections for the HL-LHC.
In all cases we have symmetrised the experimental uncertainties that we use in the fits.  }
\label{table:resHiggsExp}
\end{table}
\afterpage{\clearpage} 
 
\newpage
\section{Two parameter fits \label{App:fitplots}}

\begin{figure}[h!]
	\vspace{0 cm}
	\begin{center}
		\includegraphics[width=0.42\linewidth]{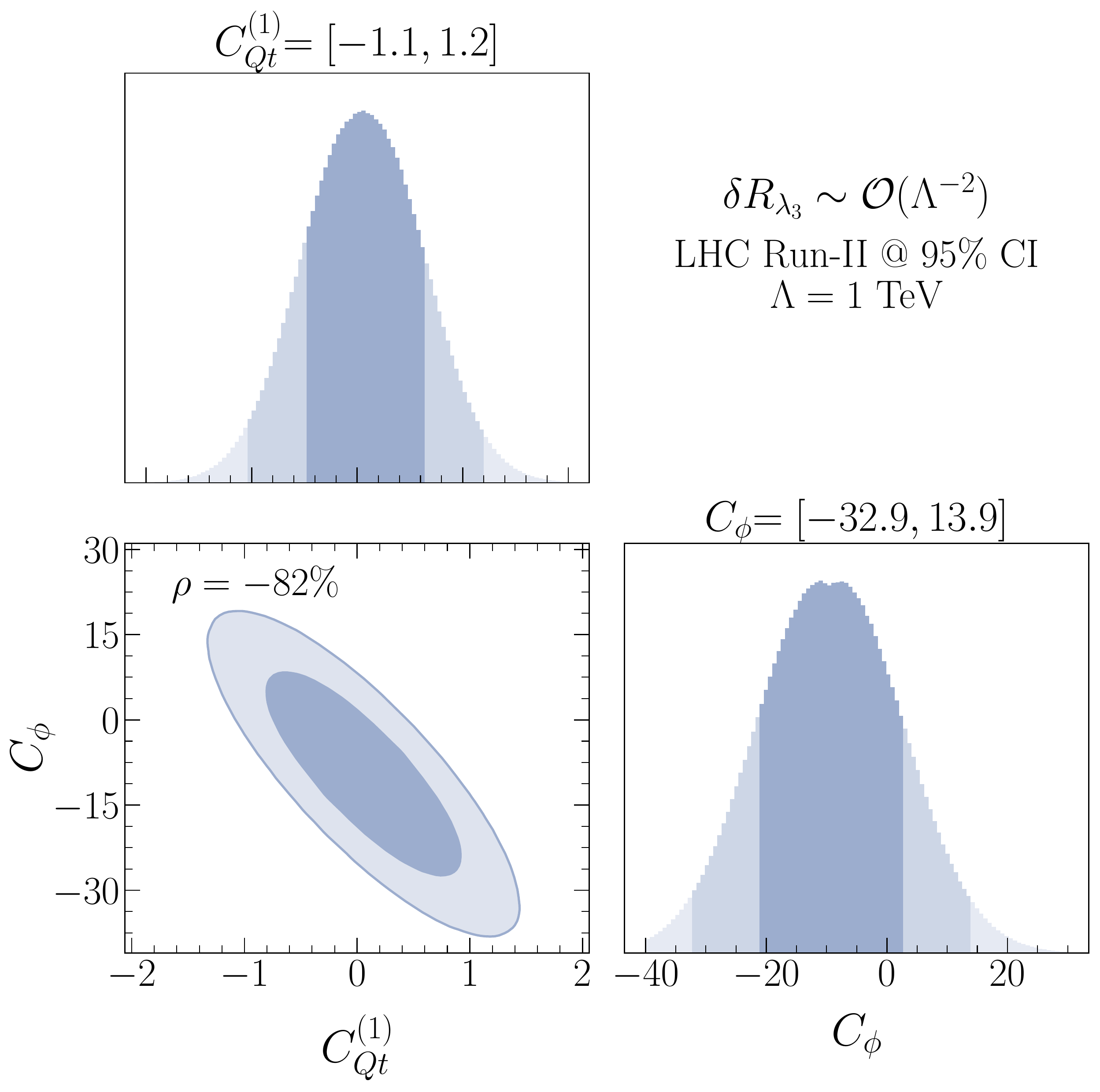}
       \includegraphics[width=0.42\linewidth]{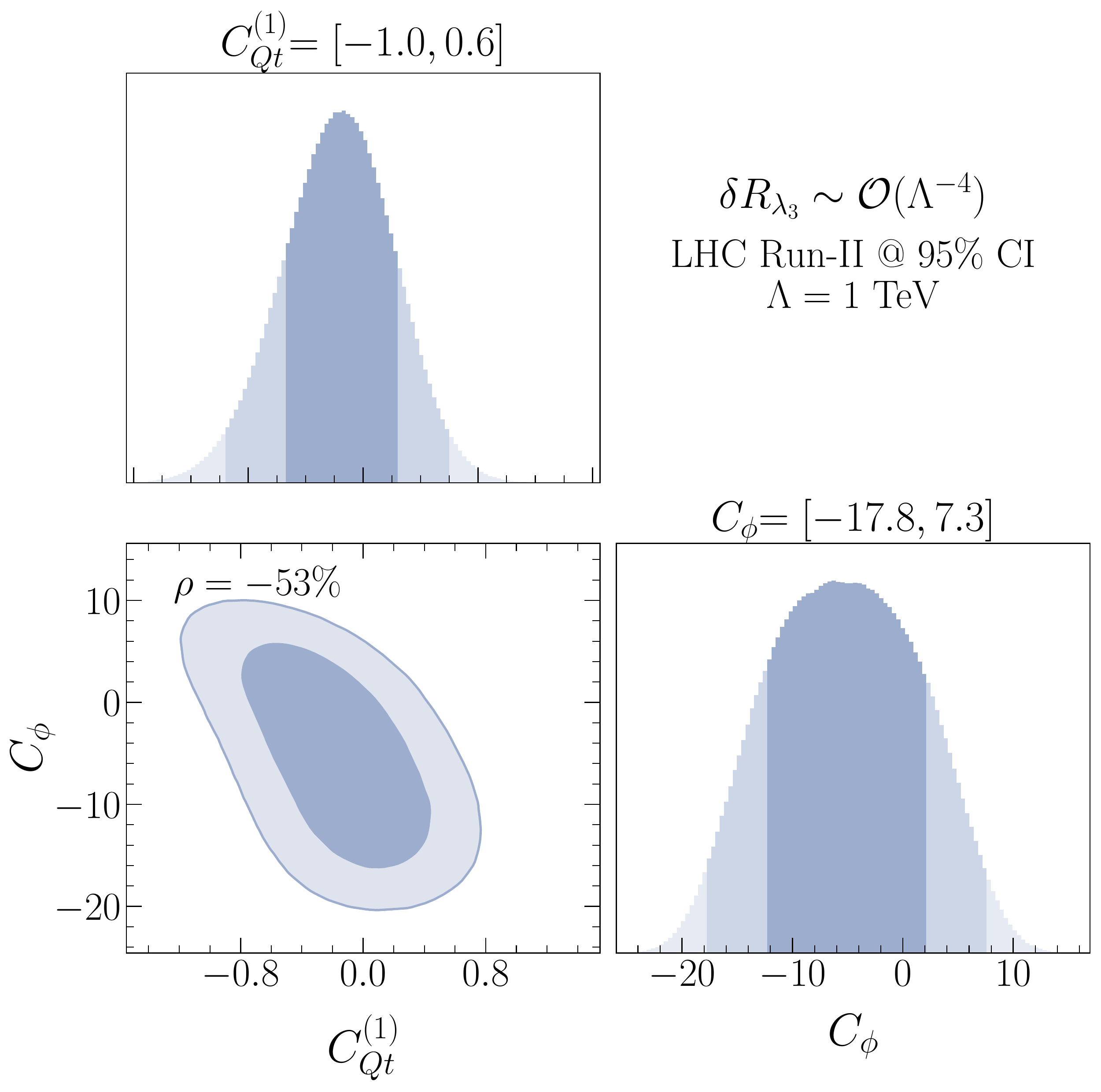} \\ 
       		\includegraphics[width=0.42\linewidth]{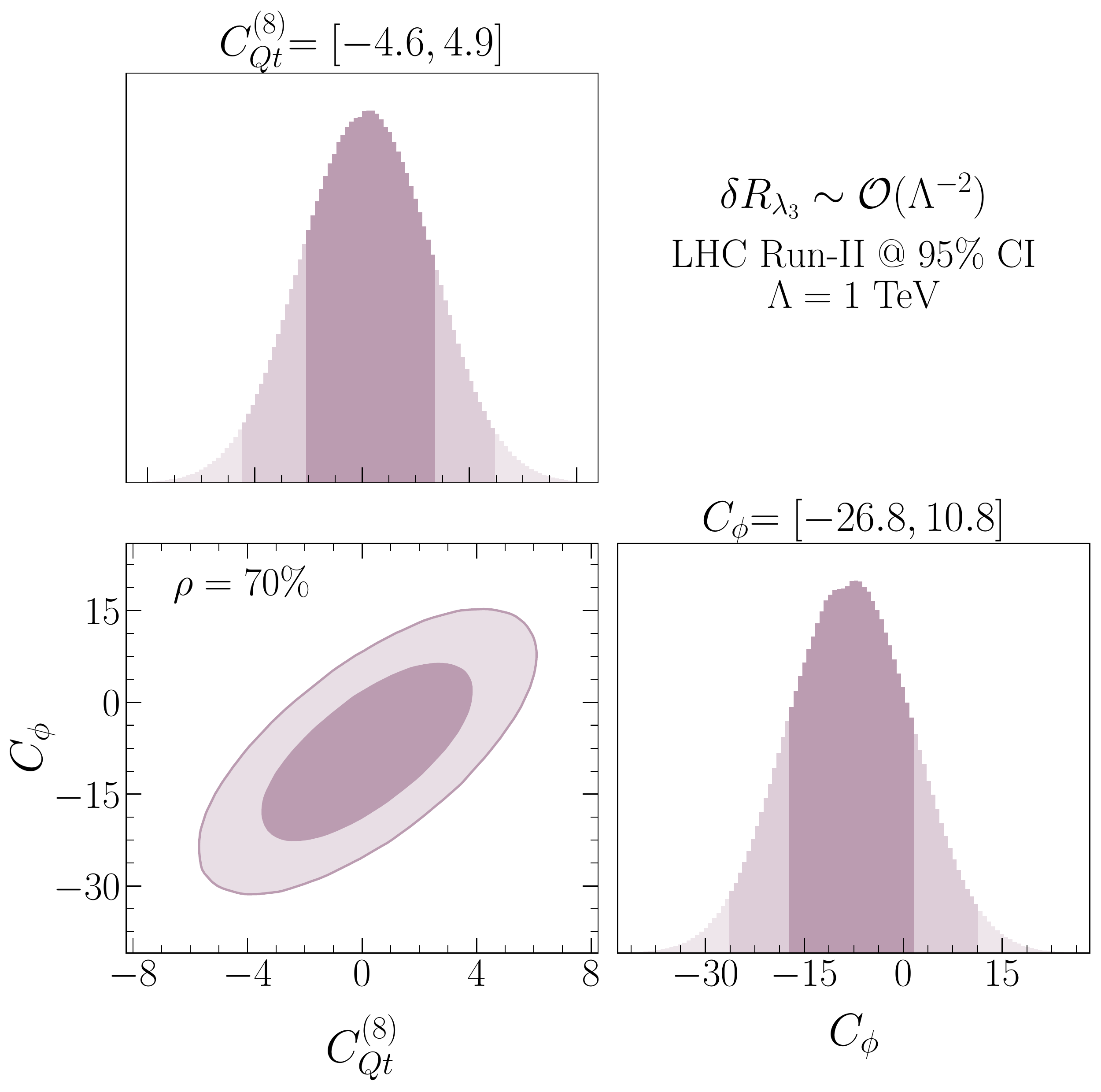}
       \includegraphics[width=0.42\linewidth]{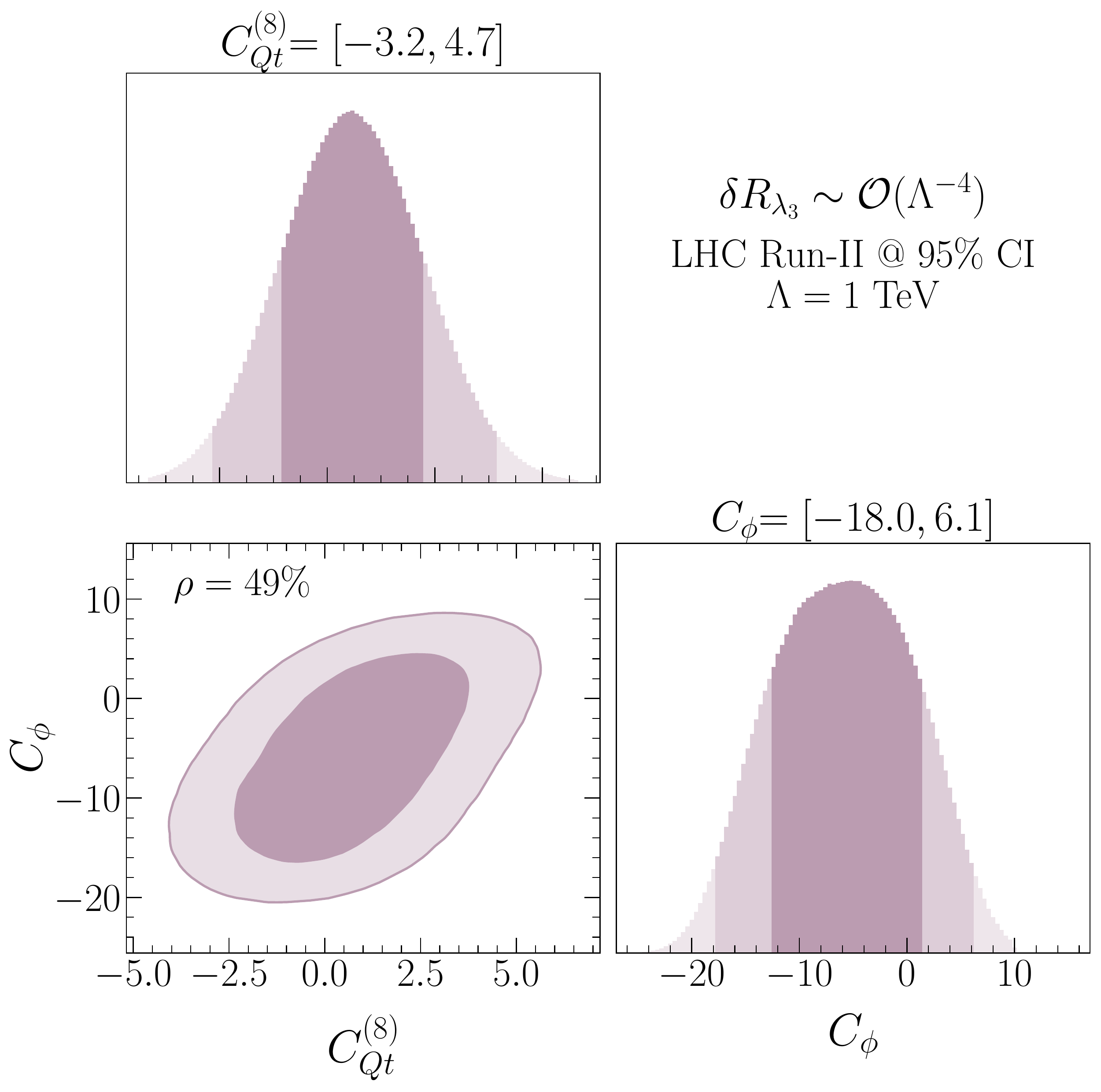} \\
	\end{center}
\vspace{-0.6 cm}
	\caption{The 68\% and 95\% highest density posterior contours of the posterior distribution from the fits of $C_\phi$ with the four-top Wilson coefficient $C_{Qt}^{(1)}$ (top panels) and $C_\phi$ with $C_{Qt}^{(8)}$  (bottom panels).  We also show the marginalised one-dimensional posteriors for each of the Wilson coefficients, with their 68\% and 95\% HDPIs and, on top of the corresponding figures, the numerical 95\% CI bounds. 
The limits correspond to values of the Wilson coefficients evaluated at the scale $\Lambda=1$ TeV.
On the left we used the linear scheme in $\delta R_{\lambda_3}$ while on the right we include effects up to $\mathcal{O}(1/\Lambda^4)$ in $\delta R_{\lambda_3}$. \label{2param-cqt} } 
\end{figure}

\begin{figure}[htpb!]
	\vspace{0 cm}
	\begin{center}
		\includegraphics[width=0.42\linewidth]{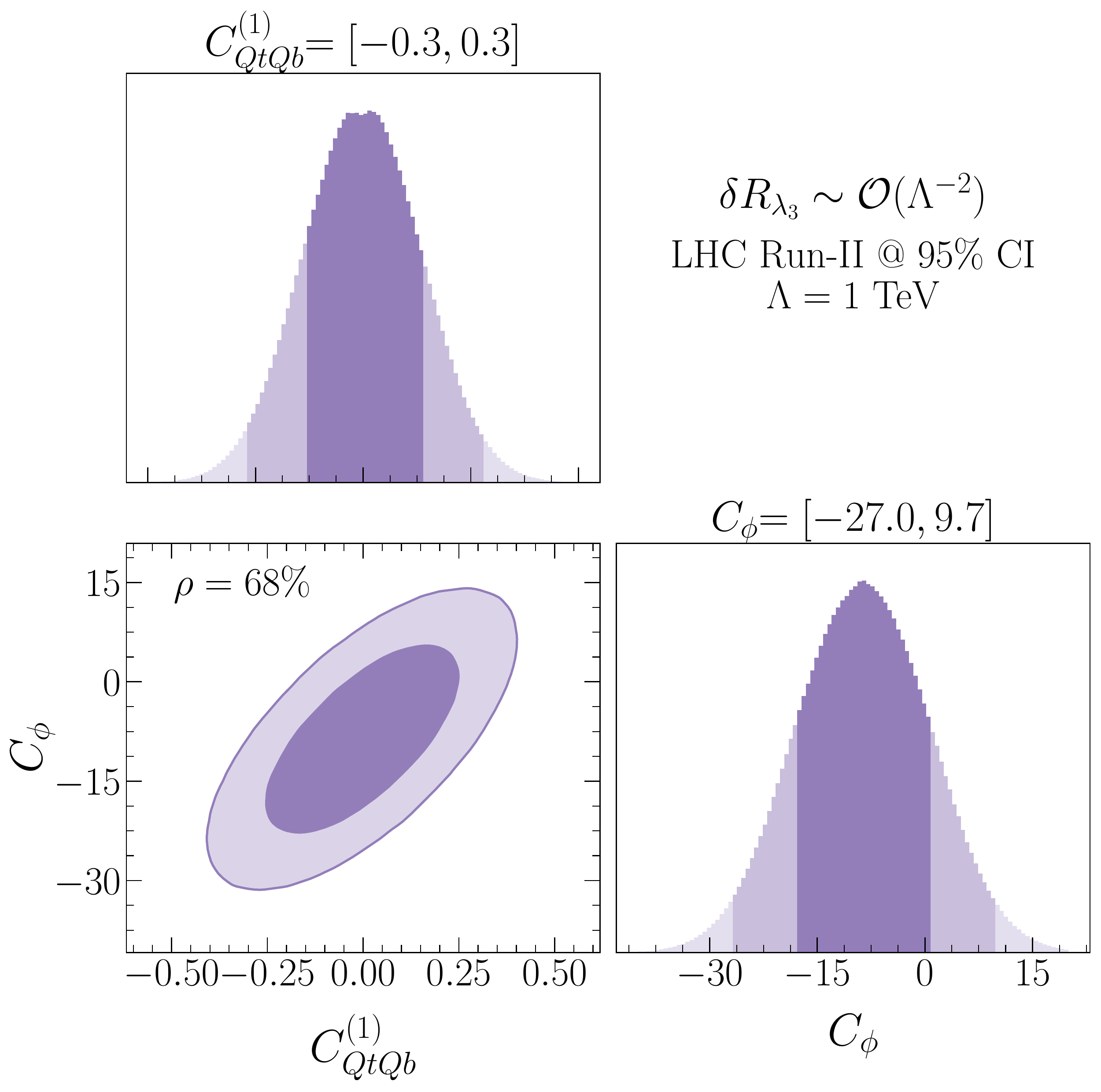}
		\includegraphics[width=0.42\linewidth]{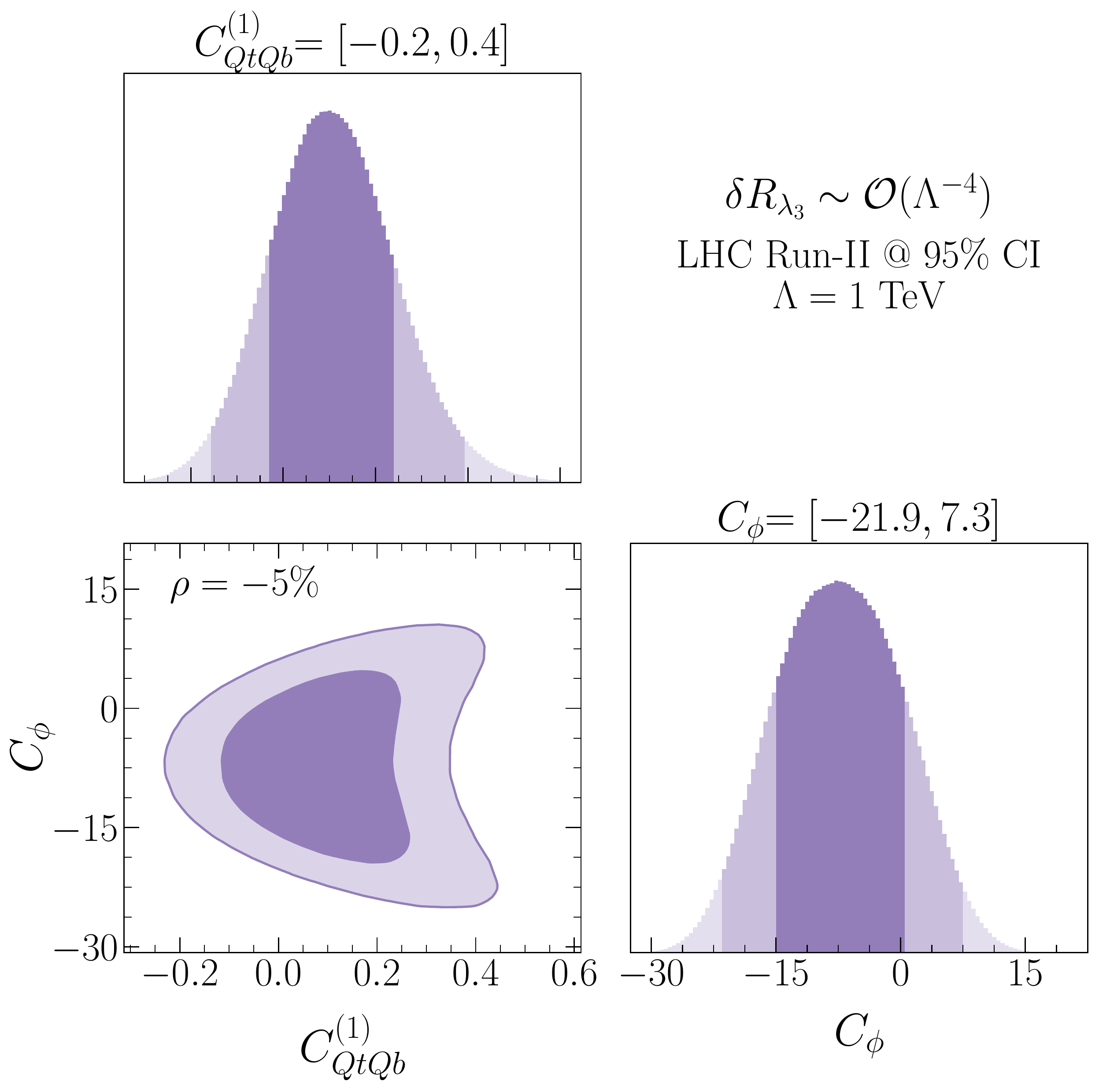} \\ 
		\includegraphics[width=0.42\linewidth]{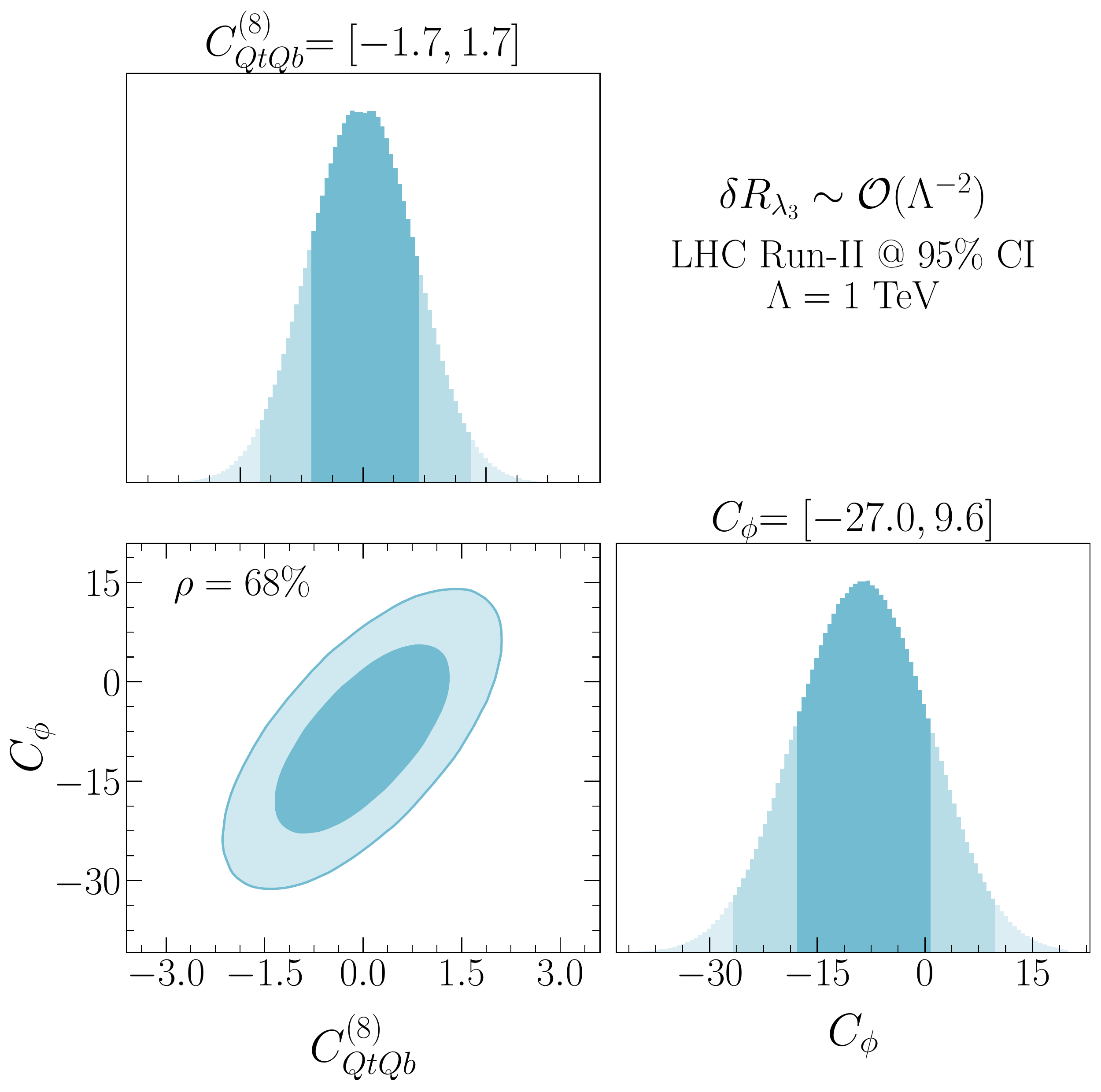}
		\includegraphics[width=0.42\linewidth]{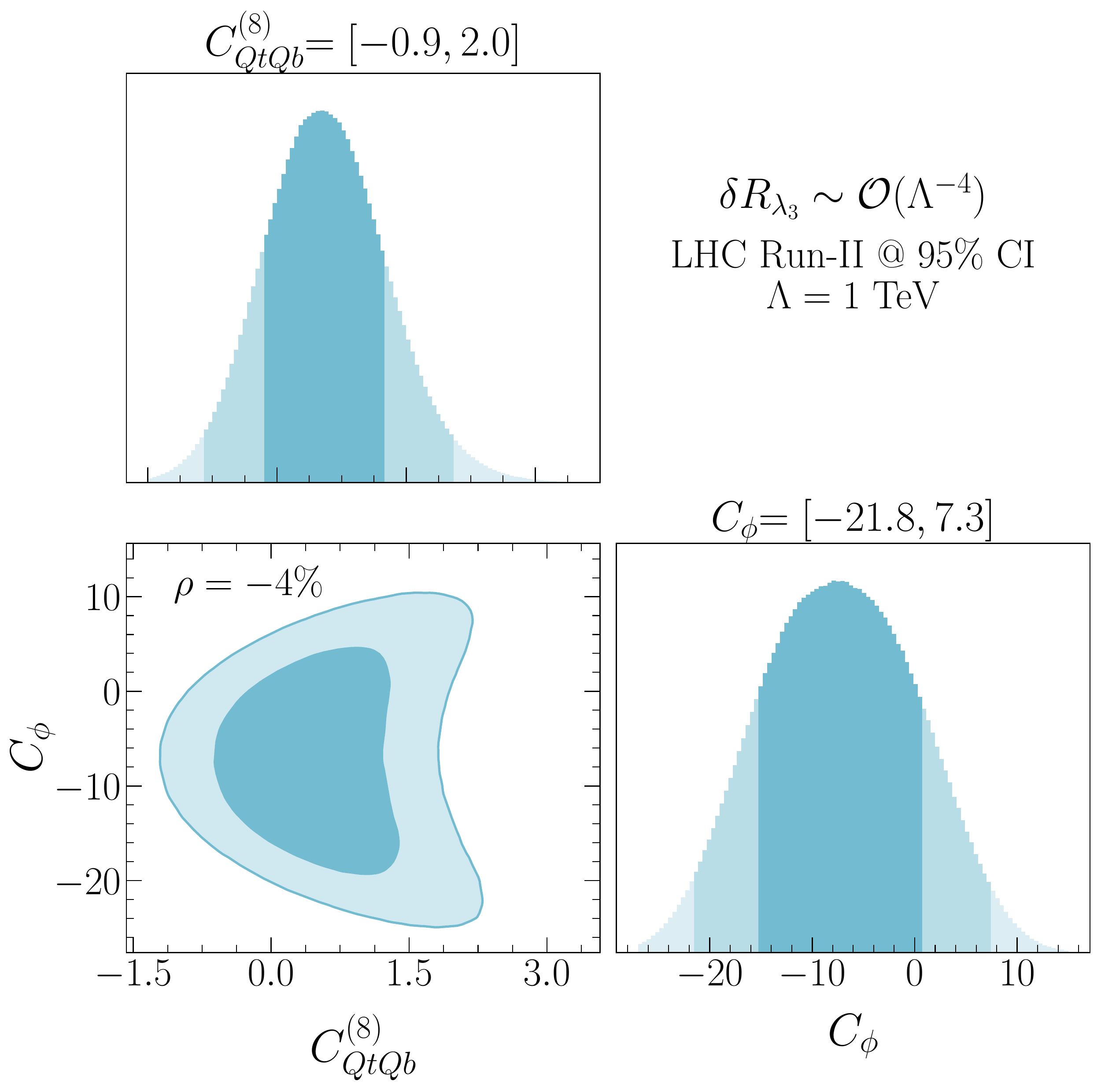} 
	\end{center}
\vspace{-0.6 cm}
	\caption{The 68\% and 95\% highest density posterior contours of the posterior distribution from the fits of $C_\phi$ with $C_{QtQb}^{(1)}$ (top panels) and $C_\phi$ with $C_{QtQb}^{(8)}$  (bottom panels).  We also show the marginalised one-dimensional posteriors for each of the Wilson coefficients, with their 68\% and 95\% HDPIs and, on top of the corresponding figures, the numerical 95\% CI bounds. 
The limits correspond to values of the Wilson coefficients evaluated at the scale $\Lambda=1$ TeV. 
		Similar to \autoref{2param-cqt}, the left plots shows the results including only $\mathcal{O}(1/\Lambda^2)$ effects in $\delta R_{\lambda_3}$, while the right ones include up to quadratic terms in the trilinear Higgs self-coupling modification. Due to the degeneracy between the Wilson coefficients $C_{QtQb}^{(1),(8)}$, the posterior contours and their marginalised intervals look very similar for both of them (except for the range they cover).  \label{2param-cqtqb} }
\end{figure}

We present in~\autoref{2param-cqt} and~\autoref{2param-cqtqb}  the $68\%$ and $ 95\%$  highest posterior density contours of the two-parameter posterior distributions and their marginalisation for the two-parameter fits involving $C_\phi$ and each of the four-quark Wilson coefficients, evaluated at the scale $\Lambda=1$ TeV.  
Both linearised and quadratically truncated $\delta R_{\lambda_3}$ fits are shown, and we observe that the $95\%$ CI bounds (shown on top of the panels) and correlations depends on the truncation.


\clearpage
\newpage

\bibliographystyle{utphys.bst}
\bibliography{bibliography}

\providecommand{\href}[2]{#2}\begingroup\raggedright\begin{thebibliography}{10}

\bibitem{Aad:2019mbh}
{\bfseries ATLAS} Collaboration, G.~Aad {\em et~al.}, ``{Combined measurements
  of Higgs boson production and decay using up to $80$ fb$^{-1}$ of
  proton-proton collision data at $\sqrt{s}=$ 13 TeV collected with the ATLAS
  experiment},'' \href{http://dx.doi.org/10.1103/PhysRevD.101.012002}{{\em
  Phys. Rev. D} {\bfseries 101} no.~1, (2020) 012002},
  \href{http://arxiv.org/abs/1909.02845}{{\ttfamily arXiv:1909.02845
  [hep-ex]}}.

\bibitem{Sirunyan:2018koj}
{\bfseries CMS} Collaboration, A.~M. Sirunyan {\em et~al.}, ``{Combined
  measurements of Higgs boson couplings in proton\textendash{}proton collisions
  at $\sqrt{s}=13\,\text {Te}\text {V} $},''
  \href{http://dx.doi.org/10.1140/epjc/s10052-019-6909-y}{{\em Eur. Phys. J. C}
  {\bfseries 79} no.~5, (2019) 421},
  \href{http://arxiv.org/abs/1809.10733}{{\ttfamily arXiv:1809.10733
  [hep-ex]}}.

\bibitem{Frederix:2017wme}
R.~Frederix, D.~Pagani, and M.~Zaro, ``{Large NLO corrections in
  $t\bar{t}W^{\pm}$ and $t\bar{t}t\bar{t}$ hadroproduction from supposedly
  subleading EW contributions},''
  \href{http://dx.doi.org/10.1007/JHEP02(2018)031}{{\em JHEP} {\bfseries 02}
  (2018) 031}, \href{http://arxiv.org/abs/1711.02116}{{\ttfamily
  arXiv:1711.02116 [hep-ph]}}.

\bibitem{Aad:2020klt}
{\bfseries ATLAS} Collaboration, G.~Aad {\em et~al.}, ``{Evidence for
  $t\bar{t}t\bar{t}$ production in the multilepton final state in
  proton\textendash{}proton collisions at $\sqrt{s}=13$ $\text {TeV}$ with the
  ATLAS detector},''
  \href{http://dx.doi.org/10.1140/epjc/s10052-020-08509-3}{{\em Eur. Phys. J.
  C} {\bfseries 80} no.~11, (2020) 1085},
  \href{http://arxiv.org/abs/2007.14858}{{\ttfamily arXiv:2007.14858
  [hep-ex]}}.

\bibitem{Sirunyan:2019nxl}
{\bfseries CMS} Collaboration, A.~M. Sirunyan {\em et~al.}, ``{Search for the
  production of four top quarks in the single-lepton and opposite-sign dilepton
  final states in proton-proton collisions at $ \sqrt{s} $ = 13 TeV},''
  \href{http://dx.doi.org/10.1007/JHEP11(2019)082}{{\em JHEP} {\bfseries 11}
  (2019) 082}, \href{http://arxiv.org/abs/1906.02805}{{\ttfamily
  arXiv:1906.02805 [hep-ex]}}.

\bibitem{Hartland:2019bjb}
N.~P. Hartland, F.~Maltoni, E.~R. Nocera, J.~Rojo, E.~Slade, E.~Vryonidou, and
  C.~Zhang, ``{A Monte Carlo global analysis of the Standard Model Effective
  Field Theory: the top quark sector},''
  \href{http://dx.doi.org/10.1007/JHEP04(2019)100}{{\em JHEP} {\bfseries 04}
  (2019) 100}, \href{http://arxiv.org/abs/1901.05965}{{\ttfamily
  arXiv:1901.05965 [hep-ph]}}.

\bibitem{Banelli:2020iau}
G.~Banelli, E.~Salvioni, J.~Serra, T.~Theil, and A.~Weiler, ``{The Present and
  Future of Four Top Operators},''
  \href{http://dx.doi.org/10.1007/JHEP02(2021)043}{{\em JHEP} {\bfseries 02}
  (2021) 043}, \href{http://arxiv.org/abs/2010.05915}{{\ttfamily
  arXiv:2010.05915 [hep-ph]}}.

\bibitem{Sirunyan:2020kgar}
{\bfseries CMS} Collaboration, A.~M. Sirunyan {\em et~al.}, ``{Measurement of
  the cross section for $\text{t}\bar{\text{t}}$ production with additional
  jets and b jets in pp collisions at $\sqrt{s}=$ 13 TeV},''
  \href{http://dx.doi.org/10.1007/JHEP07(2020)125}{{\em JHEP} {\bfseries 07}
  (2020) 125}, \href{http://arxiv.org/abs/2003.06467}{{\ttfamily
  arXiv:2003.06467 [hep-ex]}}.

\bibitem{ATLAS:2018gug}
{\bfseries ATLAS} Collaboration, ``{Measurements of fiducial and differential
  cross-sections of $t\bar{t}$ production with additional heavy-flavour jets in
  proton-proton collisions at $\sqrt{s} = 13$ TeV with the ATLAS detector},''
  Tech. Rep. ATLAS-CONF-2018-029, 2018.

\bibitem{DHondt:2018cww}
J.~D'Hondt, A.~Mariotti, K.~Mimasu, S.~Moortgat, and C.~Zhang, ``{Learning to
  pinpoint effective operators at the LHC: a study of the $
  \mathrm{t}\overline{\mathrm{t}}\mathrm{b}\overline{\mathrm{b}} $
  signature},'' \href{http://dx.doi.org/10.1007/JHEP11(2018)131}{{\em JHEP}
  {\bfseries 11} (2018) 131}, \href{http://arxiv.org/abs/1807.02130}{{\ttfamily
  arXiv:1807.02130 [hep-ph]}}.

\bibitem{Degrande:2020evl}
C.~Degrande, G.~Durieux, F.~Maltoni, K.~Mimasu, E.~Vryonidou, and C.~Zhang,
  ``{Automated one-loop computations in the standard model effective field
  theory},'' \href{http://dx.doi.org/10.1103/PhysRevD.103.096024}{{\em Phys.
  Rev. D} {\bfseries 103} no.~9, (2021) 096024},
  \href{http://arxiv.org/abs/2008.11743}{{\ttfamily arXiv:2008.11743
  [hep-ph]}}.

\bibitem{deBlas:2015aea}
J.~de~Blas, M.~Chala, and J.~Santiago, ``{Renormalization Group Constraints on
  New Top Interactions from Electroweak Precision Data},''
  \href{http://dx.doi.org/10.1007/JHEP09(2015)189}{{\em JHEP} {\bfseries 09}
  (2015) 189}, \href{http://arxiv.org/abs/1507.00757}{{\ttfamily
  arXiv:1507.00757 [hep-ph]}}.

\bibitem{Dawson:2022bxd}
S.~Dawson and P.~P. Giardino, ``{Flavorful Electroweak Precision Observables in
  the Standard Model Effective Field Theory},''
  \href{http://arxiv.org/abs/2201.09887}{{\ttfamily arXiv:2201.09887
  [hep-ph]}}.

\bibitem{Jenkins:2013zja}
E.~E. Jenkins, A.~V. Manohar, and M.~Trott, ``{Renormalization Group Evolution
  of the Standard Model Dimension Six Operators I: Formalism and lambda
  Dependence},'' \href{http://dx.doi.org/10.1007/JHEP10(2013)087}{{\em JHEP}
  {\bfseries 10} (2013) 087}, \href{http://arxiv.org/abs/1308.2627}{{\ttfamily
  arXiv:1308.2627 [hep-ph]}}.

\bibitem{Jenkins:2013wua}
E.~E. Jenkins, A.~V. Manohar, and M.~Trott, ``{Renormalization Group Evolution
  of the Standard Model Dimension Six Operators II: Yukawa Dependence},''
  \href{http://dx.doi.org/10.1007/JHEP01(2014)035}{{\em JHEP} {\bfseries 01}
  (2014) 035}, \href{http://arxiv.org/abs/1310.4838}{{\ttfamily arXiv:1310.4838
  [hep-ph]}}.

\bibitem{Alonso:2013hga}
R.~Alonso, E.~E. Jenkins, A.~V. Manohar, and M.~Trott, ``{Renormalization Group
  Evolution of the Standard Model Dimension Six Operators III: Gauge Coupling
  Dependence and Phenomenology},''
  \href{http://dx.doi.org/10.1007/JHEP04(2014)159}{{\em JHEP} {\bfseries 04}
  (2014) 159}, \href{http://arxiv.org/abs/1312.2014}{{\ttfamily arXiv:1312.2014
  [hep-ph]}}.

\bibitem{McCullough:2013rea}
M.~McCullough, ``{An Indirect Model-Dependent Probe of the Higgs
  Self-Coupling},'' \href{http://dx.doi.org/10.1103/PhysRevD.90.015001}{{\em
  Phys. Rev. D} {\bfseries 90} no.~1, (2014) 015001},
  \href{http://arxiv.org/abs/1312.3322}{{\ttfamily arXiv:1312.3322 [hep-ph]}}.
  [Erratum: Phys.Rev.D 92, 039903 (2015)].

\bibitem{Gorbahn:2016uoy}
M.~Gorbahn and U.~Haisch, ``{Indirect probes of the trilinear Higgs coupling:
  $gg \to h$ and $h \to \gamma \gamma$},''
  \href{http://dx.doi.org/10.1007/JHEP10(2016)094}{{\em JHEP} {\bfseries 10}
  (2016) 094}, \href{http://arxiv.org/abs/1607.03773}{{\ttfamily
  arXiv:1607.03773 [hep-ph]}}.

\bibitem{Degrassi:2016wml}
G.~Degrassi, P.~P. Giardino, F.~Maltoni, and D.~Pagani, ``{Probing the Higgs
  self coupling via single Higgs production at the LHC},''
  \href{http://dx.doi.org/10.1007/JHEP12(2016)080}{{\em JHEP} {\bfseries 12}
  (2016) 080}, \href{http://arxiv.org/abs/1607.04251}{{\ttfamily
  arXiv:1607.04251 [hep-ph]}}.

\bibitem{Bizon:2016wgr}
W.~Bizon, M.~Gorbahn, U.~Haisch, and G.~Zanderighi, ``{Constraints on the
  trilinear Higgs coupling from vector boson fusion and associated Higgs
  production at the LHC},''
  \href{http://dx.doi.org/10.1007/JHEP07(2017)083}{{\em JHEP} {\bfseries 07}
  (2017) 083}, \href{http://arxiv.org/abs/1610.05771}{{\ttfamily
  arXiv:1610.05771 [hep-ph]}}.

\bibitem{Maltoni:2017ims}
F.~Maltoni, D.~Pagani, A.~Shivaji, and X.~Zhao, ``{Trilinear Higgs coupling
  determination via single-Higgs differential measurements at the LHC},''
  \href{http://dx.doi.org/10.1140/epjc/s10052-017-5410-8}{{\em Eur. Phys. J. C}
  {\bfseries 77} no.~12, (2017) 887},
  \href{http://arxiv.org/abs/1709.08649}{{\ttfamily arXiv:1709.08649
  [hep-ph]}}.

\bibitem{Degrassi:2019yix}
G.~Degrassi and M.~Vitti, ``{The effect of an anomalous Higgs trilinear
  self-coupling on the $h \rightarrow \gamma \, Z$ decay},''
  \href{http://dx.doi.org/10.1140/epjc/s10052-020-7860-7}{{\em Eur. Phys. J. C}
  {\bfseries 80} no.~4, (2020) 307},
  \href{http://arxiv.org/abs/1912.06429}{{\ttfamily arXiv:1912.06429
  [hep-ph]}}.

\bibitem{Degrassi:2021uik}
G.~Degrassi, B.~Di~Micco, P.~P. Giardino, and E.~Rossi, ``{Higgs boson
  self-coupling constraints from single Higgs, double Higgs and Electroweak
  measurements},'' \href{http://dx.doi.org/10.1016/j.physletb.2021.136307}{{\em
  Phys. Lett. B} {\bfseries 817} (2021) 136307},
  \href{http://arxiv.org/abs/2102.07651}{{\ttfamily arXiv:2102.07651
  [hep-ph]}}.

\bibitem{Haisch:2021hvy}
U.~Haisch and G.~Koole, ``{Off-shell Higgs production at the LHC as a probe of
  the trilinear Higgs coupling},''
  \href{http://arxiv.org/abs/2111.12589}{{\ttfamily arXiv:2111.12589
  [hep-ph]}}.

\bibitem{DiVita:2017eyz}
S.~Di~Vita, C.~Grojean, G.~Panico, M.~Riembau, and T.~Vantalon, ``{A global
  view on the Higgs self-coupling},''
  \href{http://dx.doi.org/10.1007/JHEP09(2017)069}{{\em JHEP} {\bfseries 09}
  (2017) 069}, \href{http://arxiv.org/abs/1704.01953}{{\ttfamily
  arXiv:1704.01953 [hep-ph]}}.

\bibitem{ATLAS:2019pbo}
{\bfseries ATLAS} Collaboration, ``{Constraints on the Higgs boson
  self-coupling from the combination of single-Higgs and double-Higgs
  production analyses performed with the ATLAS experiment},'' Tech. Rep.
  ATLAS-CONF-2019-049, 2019.

\bibitem{CMS:2020gsy}
{\bfseries CMS} Collaboration, ``{Combined Higgs boson production and decay
  measurements with up to 137 fb$^{-1}$ of proton-proton collision data at
  $\sqrt s$ = 13 TeV},''.

\bibitem{Grzadkowski:2010es}
B.~Grzadkowski, M.~Iskrzynski, M.~Misiak, and J.~Rosiek, ``{Dimension-Six Terms
  in the Standard Model Lagrangian},''
  \href{http://dx.doi.org/10.1007/JHEP10(2010)085}{{\em JHEP} {\bfseries 10}
  (2010) 085}, \href{http://arxiv.org/abs/1008.4884}{{\ttfamily arXiv:1008.4884
  [hep-ph]}}.

\bibitem{Patel:2015tea}
H.~Patel, ``{Package-X: A Mathematica package for the analytic calculation of
  one-loop integrals},''
  \href{http://dx.doi.org/10.1016/j.cpc.2015.08.017}{{\em Comput. Phys.
  Commun.} {\bfseries 197} (2015) 276--290},
  \href{http://arxiv.org/abs/1503.01469}{{\ttfamily arXiv:1503.01469
  [hep-ph]}}.

\bibitem{Maierhoefer:2017hyi}
P.~Maierh{\"o}fer, J.~Usovitsch, and P.~Uwer, ``{Kira---A Feynman integral
  reduction program},'' \href{http://dx.doi.org/10.1016/j.cpc.2018.04.012}{{\em
  Comput. Phys. Commun.} {\bfseries 230} (2018) 99--112},
  \href{http://arxiv.org/abs/1705.05610}{{\ttfamily arXiv:1705.05610
  [hep-ph]}}.

\bibitem{Smirnov:2008iw}
A.~Smirnov, ``{Algorithm FIRE -- Feynman Integral REduction},''
  \href{http://dx.doi.org/10.1088/1126-6708/2008/10/107}{{\em JHEP} {\bfseries
  10} (2008) 107}, \href{http://arxiv.org/abs/0807.3243}{{\ttfamily
  arXiv:0807.3243 [hep-ph]}}.

\bibitem{Alloul:2013bka}
A.~Alloul, N.~D. Christensen, C.~Degrande, C.~Duhr, and B.~Fuks, ``{FeynRules
  2.0 - A complete toolbox for tree-level phenomenology},''
  \href{http://dx.doi.org/10.1016/j.cpc.2014.04.012}{{\em Comput. Phys.
  Commun.} {\bfseries 185} (2014) 2250--2300},
  \href{http://arxiv.org/abs/1310.1921}{{\ttfamily arXiv:1310.1921 [hep-ph]}}.

\bibitem{Hahn:2000kx}
T.~Hahn, ``{Generating Feynman diagrams and amplitudes with FeynArts 3},''
  \href{http://dx.doi.org/10.1016/S0010-4655(01)00290-9}{{\em Comput. Phys.
  Commun.} {\bfseries 140} (2001) 418--431},
  \href{http://arxiv.org/abs/hep-ph/0012260}{{\ttfamily arXiv:hep-ph/0012260}}.

\bibitem{Dedes:2017zog}
A.~Dedes, W.~Materkowska, M.~Paraskevas, J.~Rosiek, and K.~Suxho, ``{Feynman
  rules for the Standard Model Effective Field Theory in $R_{\xi}$ -gauges},''
  \href{http://dx.doi.org/10.1007/JHEP06(2017)143}{{\em JHEP} {\bfseries 06}
  (2017) 143}, \href{http://arxiv.org/abs/1704.03888}{{\ttfamily
  arXiv:1704.03888 [hep-ph]}}.

\bibitem{Dawson:2018pyl}
S.~Dawson and P.~P. Giardino, ``{Higgs decays to $ZZ$ and $Z\gamma$ in the
  standard model effective field theory: An NLO analysis},''
  \href{http://dx.doi.org/10.1103/PhysRevD.97.093003}{{\em Phys. Rev. D}
  {\bfseries 97} no.~9, (2018) 093003},
  \href{http://arxiv.org/abs/1801.01136}{{\ttfamily arXiv:1801.01136
  [hep-ph]}}.

\bibitem{Gauld:2015lmb}
R.~Gauld, B.~D. Pecjak, and D.~J. Scott, ``{One-loop corrections to $h\to b\bar
  b$ and $h\to \tau\bar \tau$ decays in the Standard Model Dimension-6 EFT:
  four-fermion operators and the large-$m_t$ limit},''
  \href{http://dx.doi.org/10.1007/JHEP05(2016)080}{{\em JHEP} {\bfseries 05}
  (2016) 080}, \href{http://arxiv.org/abs/1512.02508}{{\ttfamily
  arXiv:1512.02508 [hep-ph]}}.

\bibitem{Ossola:2006us}
G.~Ossola, C.~G. Papadopoulos, and R.~Pittau, ``{Reducing full one-loop
  amplitudes to scalar integrals at the integrand level},''
  \href{http://dx.doi.org/10.1016/j.nuclphysb.2006.11.012}{{\em Nucl. Phys. B}
  {\bfseries 763} (2007) 147--169},
  \href{http://arxiv.org/abs/hep-ph/0609007}{{\ttfamily arXiv:hep-ph/0609007}}.

\bibitem{Ossola:2007ax}
G.~Ossola, C.~G. Papadopoulos, and R.~Pittau, ``{CutTools: A Program
  implementing the OPP reduction method to compute one-loop amplitudes},''
  \href{http://dx.doi.org/10.1088/1126-6708/2008/03/042}{{\em JHEP} {\bfseries
  03} (2008) 042}, \href{http://arxiv.org/abs/0711.3596}{{\ttfamily
  arXiv:0711.3596 [hep-ph]}}.

\bibitem{Ossola:2008xq}
G.~Ossola, C.~G. Papadopoulos, and R.~Pittau, ``{On the Rational Terms of the
  one-loop amplitudes},''
  \href{http://dx.doi.org/10.1088/1126-6708/2008/05/004}{{\em JHEP} {\bfseries
  05} (2008) 004}, \href{http://arxiv.org/abs/0802.1876}{{\ttfamily
  arXiv:0802.1876 [hep-ph]}}.

\bibitem{Alwall:2014hca}
J.~Alwall, R.~Frederix, S.~Frixione, V.~Hirschi, F.~Maltoni, O.~Mattelaer,
  H.~S. Shao, T.~Stelzer, P.~Torrielli, and M.~Zaro, ``{The automated
  computation of tree-level and next-to-leading order differential cross
  sections, and their matching to parton shower simulations},''
  \href{http://dx.doi.org/10.1007/JHEP07(2014)079}{{\em JHEP} {\bfseries 07}
  (2014) 079}, \href{http://arxiv.org/abs/1405.0301}{{\ttfamily arXiv:1405.0301
  [hep-ph]}}.

\bibitem{Chetyrkin:2000yt}
K.~G. Chetyrkin, J.~H. Kuhn, and M.~Steinhauser, ``{RunDec: A Mathematica
  package for running and decoupling of the strong coupling and quark
  masses},'' \href{http://dx.doi.org/10.1016/S0010-4655(00)00155-7}{{\em
  Comput. Phys. Commun.} {\bfseries 133} (2000) 43--65},
  \href{http://arxiv.org/abs/hep-ph/0004189}{{\ttfamily arXiv:hep-ph/0004189}}.

\bibitem{Zyla:2020zbs}
{\bfseries Particle Data Group} Collaboration, P.~Zyla {\em et~al.}, ``{Review
  of Particle Physics},'' \href{http://dx.doi.org/10.1093/ptep/ptaa104}{{\em
  PTEP} {\bfseries 2020} no.~8, (2020) 083C01}.

\bibitem{Ball:2012cx}
R.~D. Ball {\em et~al.}, ``{Parton distributions with LHC data},''
  \href{http://dx.doi.org/10.1016/j.nuclphysb.2012.10.003}{{\em Nucl. Phys. B}
  {\bfseries 867} (2013) 244--289},
  \href{http://arxiv.org/abs/1207.1303}{{\ttfamily arXiv:1207.1303 [hep-ph]}}.

\bibitem{LHCHiggsCrossSectionWorkingGroup:2016ypw}
{\bfseries LHC Higgs Cross Section Working Group} Collaboration, D.~de~Florian
  {\em et~al.}, ``{Handbook of LHC Higgs Cross Sections: 4. Deciphering the
  Nature of the Higgs Sector},''
  \href{http://arxiv.org/abs/1610.07922}{{\ttfamily arXiv:1610.07922
  [hep-ph]}}.

\bibitem{Salvatier2016}
J.~Salvatier, T.~V. Wiecki, and C.~Fonnesbeck, ``Probabilistic programming in
  python using {PyMC}3,'' \href{http://dx.doi.org/10.7717/peerj-cs.55}{{\em
  {PeerJ} Computer Science} {\bfseries 2} (Apr, 2016) e55}.
  \url{https://doi.org/10.7717/peerj-cs.55}.

\bibitem{arviz_2019}
R.~Kumar, C.~Carroll, A.~Hartikainen, and O.~Martin, ``Arviz a unified library
  for exploratory analysis of bayesian models in python,''
  \href{http://dx.doi.org/10.21105/joss.01143}{{\em Journal of Open Source
  Software} {\bfseries 4} no.~33, (2019) 1143}.
  \url{https://doi.org/10.21105/joss.01143}.

\bibitem{James:1975dr}
F.~James and M.~Roos, ``{Minuit: A System for Function Minimization and
  Analysis of the Parameter Errors and Correlations},''
  \href{http://dx.doi.org/10.1016/0010-4655(75)90039-9}{{\em Comput. Phys.
  Commun.} {\bfseries 10} (1975) 343--367}.

\bibitem{iminuit}
H.~Dembinski and P.~O. et~al., ``scikit-hep/iminuit,''.
  \url{https://doi.org/10.5281/zenodo.3949207}.

\bibitem{corner}
D.~Foreman-Mackey, ``corner.py: Scatterplot matrices in python,''
  \href{http://dx.doi.org/10.21105/joss.00024}{{\em The Journal of Open Source
  Software} {\bfseries 1} no.~2, (Jun, 2016) 24}.
  \url{https://doi.org/10.21105/joss.00024}.

\bibitem{Bocquet2016}
S.~Bocquet and F.~W. Carter, ``pygtc: beautiful parameter covariance plots
  (aka. giant triangle confusograms),''
  \href{http://dx.doi.org/10.21105/joss.00046}{{\em The Journal of Open Source
  Software} {\bfseries 1} no.~6, (Oct, 2016) }.
  \url{http://dx.doi.org/10.21105/joss.00046}.

\bibitem{paul_zivich_2019_3339870}
P.~Zivich, ``{pzivich/Python-for-Epidemiologists: Updates for v0.8.0},'' July,
  2019.
\newblock \url{https://doi.org/10.5281/zenodo.3339870}.

\bibitem{GitHub}
\url{https://github.com/alasfar-lina/trilinear4tops}.

\bibitem{DeBlas:2019ehy}
J.~de~Blas {\em et~al.}, ``{$\texttt{HEPfit}$: a code for the combination of
  indirect and direct constraints on high energy physics models},''
  \href{http://dx.doi.org/10.1140/epjc/s10052-020-7904-z}{{\em Eur. Phys. J. C}
  {\bfseries 80} no.~5, (2020) 456},
  \href{http://arxiv.org/abs/1910.14012}{{\ttfamily arXiv:1910.14012
  [hep-ph]}}.

\bibitem{Ethier:2021bye}
{\bfseries SMEFiT} Collaboration, J.~J. Ethier, G.~Magni, F.~Maltoni,
  L.~Mantani, E.~R. Nocera, J.~Rojo, E.~Slade, E.~Vryonidou, and C.~Zhang,
  ``{Combined SMEFT interpretation of Higgs, diboson, and top quark data from
  the LHC},'' \href{http://dx.doi.org/10.1007/JHEP11(2021)089}{{\em JHEP}
  {\bfseries 11} (2021) 089}, \href{http://arxiv.org/abs/2105.00006}{{\ttfamily
  arXiv:2105.00006 [hep-ph]}}.

\bibitem{Ellis:2020unq}
J.~Ellis, M.~Madigan, K.~Mimasu, V.~Sanz, and T.~You, ``{Top, Higgs, Diboson
  and Electroweak Fit to the Standard Model Effective Field Theory},''
  \href{http://dx.doi.org/10.1007/JHEP04(2021)279}{{\em JHEP} {\bfseries 04}
  (2021) 279}, \href{http://arxiv.org/abs/2012.02779}{{\ttfamily
  arXiv:2012.02779 [hep-ph]}}.

\bibitem{Brivio:2019ius}
I.~Brivio, S.~Bruggisser, F.~Maltoni, R.~Moutafis, T.~Plehn, E.~Vryonidou,
  S.~Westhoff, and C.~Zhang, ``{O new physics, where art thou? A global search
  in the top sector},'' \href{http://dx.doi.org/10.1007/JHEP02(2020)131}{{\em
  JHEP} {\bfseries 02} (2020) 131},
  \href{http://arxiv.org/abs/1910.03606}{{\ttfamily arXiv:1910.03606
  [hep-ph]}}.

\bibitem{Zhang:2017mls}
C.~Zhang, ``{Constraining $qqtt$ operators from four-top production: a case for
  enhanced EFT sensitivity},''
  \href{http://dx.doi.org/10.1088/1674-1137/42/2/023104}{{\em Chin. Phys. C}
  {\bfseries 42} no.~2, (2018) 023104},
  \href{http://arxiv.org/abs/1708.05928}{{\ttfamily arXiv:1708.05928
  [hep-ph]}}.

\bibitem{ATLAS:2021jki}
{\bfseries ATLAS} Collaboration, ``{Search for Higgs boson pair production in
  the two bottom quarks plus two photons final state in $pp$ collisions at
  $\sqrt{s}=13$ TeV with the ATLAS detector},'' Tech. Rep. ATLAS-CONF-2021-016,
  2021.

\bibitem{DiLuzio:2017tfn}
L.~Di~Luzio, R.~Gr\"ober, and M.~Spannowsky, ``{Maxi-sizing the trilinear Higgs
  self-coupling: how large could it be?},''
  \href{http://dx.doi.org/10.1140/epjc/s10052-017-5361-0}{{\em Eur. Phys. J. C}
  {\bfseries 77} no.~11, (2017) 788},
  \href{http://arxiv.org/abs/1704.02311}{{\ttfamily arXiv:1704.02311
  [hep-ph]}}.

\bibitem{IML}
L.~Alasfar, R.~Gr\"ober, C.~Grojean, A.~Paul, and Z.~Qian, ``{Machine learning
  augmented probes of light-quark Yukawa and trilinear couplings from Higgs
  pair production},'' {\em In preparation} (2021) .

\bibitem{CMS-PAS-FTR-18-011}
{\bfseries CMS Collaboration} Collaboration, ``{Sensitivity projections for
  Higgs boson properties measurements at the HL-LHC},'' tech. rep., CERN,
  Geneva, 2018.
\newblock \url{https://cds.cern.ch/record/2647699}.

\bibitem{twiki}
``Guidelines for extrapolation of cms and atlas lhc/hl-lhc couplings
  projections to he-lhc.''
\newblock
  \url{https://twiki.cern.ch/twiki/bin/view/LHCPhysics/GuidelinesCouplingProjections2018#Details%20of%20the%20CMS%20projections}.

\bibitem{Cepeda:2019klc}
M.~Cepeda {\em et~al.}, ``{Report from Working Group 2}: {Higgs Physics at the
  HL-LHC and HE-LHC},''
  \href{http://dx.doi.org/10.23731/CYRM-2019-007.221}{{\em CERN Yellow Rep.
  Monogr.} {\bfseries 7} (2019) 221--584},
  \href{http://arxiv.org/abs/1902.00134}{{\ttfamily arXiv:1902.00134
  [hep-ph]}}.

\bibitem{DiMicco:2019ngk}
J.~Alison {\em et~al.}, ``{Higgs boson potential at colliders: Status and
  perspectives},'' \href{http://dx.doi.org/10.1016/j.revip.2020.100045}{{\em
  Rev. Phys.} {\bfseries 5} (2020) 100045},
  \href{http://arxiv.org/abs/1910.00012}{{\ttfamily arXiv:1910.00012
  [hep-ph]}}.

\bibitem{CMS:2018ccd}
{\bfseries CMS} Collaboration, ``{Prospects for HH measurements at the
  HL-LHC},'' {\em CMS-PAS-FTR-18-019} (2018) .

\bibitem{FCC:2018byv}
{\bfseries FCC} Collaboration, A.~Abada {\em et~al.}, ``{FCC Physics
  Opportunities}: {Future Circular Collider Conceptual Design Report Volume
  1},'' \href{http://dx.doi.org/10.1140/epjc/s10052-019-6904-3}{{\em Eur. Phys.
  J. C} {\bfseries 79} no.~6, (2019) 474}.

\bibitem{FCC:2018evy}
{\bfseries FCC} Collaboration, A.~Abada {\em et~al.}, ``{FCC-ee: The Lepton
  Collider}: {Future Circular Collider Conceptual Design Report Volume 2},''
  \href{http://dx.doi.org/10.1140/epjst/e2019-900045-4}{{\em Eur. Phys. J. ST}
  {\bfseries 228} no.~2, (2019) 261--623}.

\bibitem{Bambade:2019fyw}
P.~Bambade {\em et~al.}, ``{The International Linear Collider: A Global
  Project},'' \href{http://arxiv.org/abs/1903.01629}{{\ttfamily
  arXiv:1903.01629 [hep-ex]}}.

\bibitem{LCCPhysicsWorkingGroup:2019fvj}
{\bfseries LCC Physics Working Group} Collaboration, K.~Fujii {\em et~al.},
  ``{Tests of the Standard Model at the International Linear Collider},''
  \href{http://arxiv.org/abs/1908.11299}{{\ttfamily arXiv:1908.11299
  [hep-ex]}}.

\bibitem{An:2018dwb}
F.~An {\em et~al.}, ``{Precision Higgs physics at the CEPC},''
  \href{http://dx.doi.org/10.1088/1674-1137/43/4/043002}{{\em Chin. Phys. C}
  {\bfseries 43} no.~4, (2019) 043002},
  \href{http://arxiv.org/abs/1810.09037}{{\ttfamily arXiv:1810.09037
  [hep-ex]}}.

\bibitem{CEPCStudyGroup:2018ghi}
{\bfseries CEPC Study Group} Collaboration, M.~Dong {\em et~al.}, ``{CEPC
  Conceptual Design Report: Volume 2 - Physics \& Detector},''
  \href{http://arxiv.org/abs/1811.10545}{{\ttfamily arXiv:1811.10545
  [hep-ex]}}.

\bibitem{CLICdp:2018cto}
{\bfseries CLICdp, CLIC} Collaboration, T.~K. Charles {\em et~al.}, ``{The
  Compact Linear Collider (CLIC) - 2018 Summary Report},''
  \href{http://arxiv.org/abs/1812.06018}{{\ttfamily arXiv:1812.06018
  [physics.acc-ph]}}.

\bibitem{deBlas:2018mhx}
J.~de~Blas {\em et~al.}, ``{The CLIC Potential for New Physics},''
  \href{http://arxiv.org/abs/1812.02093}{{\ttfamily arXiv:1812.02093
  [hep-ph]}}.

\bibitem{DiVita:2017vrr}
S.~Di~Vita, G.~Durieux, C.~Grojean, J.~Gu, Z.~Liu, G.~Panico, M.~Riembau, and
  T.~Vantalon, ``{A global view on the Higgs self-coupling at lepton
  colliders},'' \href{http://dx.doi.org/10.1007/JHEP02(2018)178}{{\em JHEP}
  {\bfseries 02} (2018) 178}, \href{http://arxiv.org/abs/1711.03978}{{\ttfamily
  arXiv:1711.03978 [hep-ph]}}.

\bibitem{deBlas:2017xtg}
J.~de~Blas, J.~C. Criado, M.~Perez-Victoria, and J.~Santiago, ``{Effective
  description of general extensions of the Standard Model: the complete
  tree-level dictionary},''
  \href{http://dx.doi.org/10.1007/JHEP03(2018)109}{{\em JHEP} {\bfseries 03}
  (2018) 109}, \href{http://arxiv.org/abs/1711.10391}{{\ttfamily
  arXiv:1711.10391 [hep-ph]}}.

\bibitem{Anisha:2021hgc}
Anisha, S.~D. Bakshi, S.~Banerjee, A.~Biek\"otter, J.~Chakrabortty, S.~K.
  Patra, and M.~Spannowsky, ``{Effective limits on single scalar extensions in
  the light of recent LHC data},''
  \href{http://arxiv.org/abs/2111.05876}{{\ttfamily arXiv:2111.05876
  [hep-ph]}}.

\bibitem{Grojean:2020ech}
C.~Grojean, A.~Paul, and Z.~Qian, ``{Resurrecting $ b\overline{b}h $ with
  kinematic shapes},'' \href{http://dx.doi.org/10.1007/JHEP04(2021)139}{{\em
  JHEP} {\bfseries 04} (2021) 139},
  \href{http://arxiv.org/abs/2011.13945}{{\ttfamily arXiv:2011.13945
  [hep-ph]}}.

\bibitem{ATLAS:2020qdt}
{\bfseries ATLAS} Collaboration, ``{A combination of measurements of Higgs
  boson production and decay using up to $139$ fb$^{-1}$ of proton--proton
  collision data at $\sqrt{s}=$ 13 TeV collected with the ATLAS experiment},''
  Tech. Rep. ATLAS-CONF-2020-027, 2020.

\bibitem{CMS:2021kom}
{\bfseries CMS} Collaboration, A.~M. Sirunyan {\em et~al.}, ``{Measurements of
  Higgs boson production cross sections and couplings in the diphoton decay
  channel at $ \sqrt{\mathrm{s}} $ = 13 TeV},''
  \href{http://dx.doi.org/10.1007/JHEP07(2021)027}{{\em JHEP} {\bfseries 07}
  (2021) 027}, \href{http://arxiv.org/abs/2103.06956}{{\ttfamily
  arXiv:2103.06956 [hep-ex]}}.

\bibitem{CMS:2021ixs}
{\bfseries CMS} Collaboration, ``{Measurement of Higgs boson production in
  association with a W or Z boson in the H $\rightarrow$ WW decay channel},''
  Tech. Rep. CMS-PAS-HIG-19-017, 2021.

\end{thebibliography}\endgroup

\end{document}